\documentclass{article}
\usepackage{framed}
\usepackage{graphicx} 
\usepackage{latexsym}
\usepackage{appendix}
\usepackage{hyperref}
\usepackage{setspace}
\usepackage{biblatex}
\usepackage{caption}
\newcommand{\companion}{Companion}
\title{\textbf{An Embodied Companion \\ for Visual Storytelling}}
\author{
  Patrick Tresset\\
  APT\footnote{{\href{https://patricktresset.com}{Ateliers Patrick Tresset SRL}}}\\
  Goldsmiths, University of London\\
  research@patricktresset.com
  \and
  Markus Wulfmeier\\
  Google DeepMind\\
  mwulfmeier@google.com
}
\date{}
\begin{document}

\maketitle
\begin{abstract}
 As artificial intelligence shifts from pure tool for delegation toward agentic collaboration, its use in the arts can shift beyond the exploration of machine autonomy toward synergistic co-creation. While our earlier robotic works utilized automation to distance the artist’s intent from the final mark \cite{tresset_portrait_2013}, we present \companion: an artistic apparatus that integrates a drawing robot with Large Language Models (LLMs) to re-center human-machine presence. By leveraging in-context learning and real-time tool use, the system engages in bidirectional interaction via speech and sketching. This approach transforms the robot from a passive executor into a playful co-creative partner capable of driving shared visual storytelling into unexpected aesthetic territories. To validate this collaborative shift, we employed the Consensual Assessment Technique (CAT) with a panel of seven art-world experts. Results confirm that the system produces works with a distinct aesthetic identity and professional exhibition merit, demonstrating the potential of AI as a highly capable artistic collaborator.
\footnote{We present videos of the system interacting with the artist \href{https://patricktresset.com/new/companion/}{here (https://patricktresset.com/new/companion/)}}
\end{abstract}
\section{Introduction}

The use of embodied artificial intelligence (AI) in artistic practices dates back to Harold Cohen and his AARON project in the 1970s \cite{cohen_purpose_2022}. Since then, artists and researchers have explored using robots to produce artworks, including Pindar Van Arnam, Sougwen Chung, Liat Grayver, Leonel Moura, Jon McCormack, and O. Deussen, among others \cite{grba_transparency_2023, chung_sketching_2022,grayver_transhuman_2019,moura_machines_2016, tresset_artistically_2014}. More recently, with digital art gaining acceptance in the art market, artists use robots and pen plotters to produce works based directly on digital imagery, often motivated by the perceptual engagement offered by physical artworks. 

We have been exploring the use of embodied AI systems since 2009, initially focusing on observational drawing and performance \cite{tresset_portrait_2013, tresset_artistically_2014}. 
Our motivation for using embodied computational systems was to remove ourselves from the artwork's execution.  In developing systems that could autonomously react to visual stimuli by producing drawings, our presence remained, but only in the intentions, concepts, and implementation. Increasingly capable artificial intelligence systems, such as Google's Gemini \cite{team_gemini_2025, geminiroboticsteam2025geminirobotics15pushing}, have evolved beyond providing autonomous intelligence to acting as sophisticated collaborators to work alongside humans in both digital and physical environments.

Within the context of the 'Embodied Agents in Contemporary Visual Art" (EACVA) project led by the University of Konstanz and Goldsmiths, University of London\footnote{www.eacva.org}—a multidisciplinary initiative launched in 2024 involving artists, engineers, philosophers, and psychologists —  we have been exploring the possibility of reintroducing the artist's hand during the execution process. The outcome of this research is an Embodied \companion ~for Storytelling Exploration. Together, we draw narratives inspired by children's drawings, prehistoric cave paintings, and other cultural influences.

This paper presents the development and evaluation of the \companion's prototype. The cooperative process involves the artist interacting with the agent with speech and direct physical interventions on the drawing, facilitating a visual and verbal dialogue. Our early results using the Gemini series of models demonstrate the feasibility of leveraging LLMs (hereafter referred to simply as LLMs), In-Context Learning (ICL), tools, and conversational interaction to develop a capable, embodied drawing partner \cite{team_gemini_2025}.
The following Section \ref{motivation} presents the artistic motivations for integrating drawing, storytelling, and LLMs in an embodied AI system. The remainder of this paper is structured as follows: Section \ref{related-works} presents both artistic and research-related works. Section \ref{method} presents our strategy to develop the prototype and details the system architecture and implementation. Section \ref{experiments} presents our experiments and observations. Section \ref{sec:evaluation} details the CAT study, and Section \ref{discussion} summarises the experimental outcome and presents our findings. Lastly, we conclude with a short summary and present future directions for our research.

\section{Motivation}\label{motivation}

After fifteen years of working with robots that draw on their own, a conversation with a curator pointed out a major gap in the work: the absence of the artist. Our original goal was to step back and see what the machine could do by itself. However, we found that while total autonomy produces drawings that look consistent, they often lack the meaning and feeling that come from real human experience. Early attempts to simply draw on top of robot-generated images felt fake and initial cooperative experiments for the EACVA project—repeating standard subjects like portraits and still lifes—lacked motivation. We realized that to find true engagement, we needed to recapture the "childish excitement" of drawing stories.
Consequently, the primary motivation behind this work is to shift from detached control to playful involvement, using an AI-based system to explore narrative style-spaces \cite{tresset_artistically_2014}. While artists seek the "shock of the new," they often fall into habitual patterns; we aim to counter this by utilizing the system as a partner on the journey. By allowing the AI to pull the artist in unexpected directions, we treat sketching as an adventure, opening up new territories and re-establishing the artist’s direct presence in the process.
Our choice of embodied robotics and physical drawing over purely digital methods, such as text-to-image models, is central to our artistic practice. We allow the characteristics of the physical system, its mechanics, actuators, sensors, the friction of pen on paper, as well as its computational characteristics, to shape the artwork's style. Often, the effect of these combinations is perceived as randomness or imprecision, but instead, they are factors shaping an emergent style. The lines produced are a memory of the computational and physical forces at play, a record of the process, rather than just the replication of a digital image. Furthermore, drawing unfolds in time and space, offering a distinct mode of engagement. This process evokes an immediacy linked to the directness of mark-making.  This observation may engage neural mechanisms related to action understanding, triggering a connection to the artist's actions and intentions \cite{pignocchi_how_2010}.
Inspired by the narrative qualities of children's art, the turn toward visual storytelling provides a framework for playful exploration. These forms often use simple visual elements and recurring motifs to create non-linear mini-narratives that depict actions and relationships. \companion ~aims to facilitate these types of visual explorations.

Using LLMs provides the intelligence layer for interpreting the artist's verbal and visual communications and responding to them with meaningful drawing or verbal actions. Moreover, the sequential, context-dependent nature inherent in both drawing and storytelling aligns with the strengths of transformer architectures in processing sequences and managing contextual relationships. We hypothesise that, as style emerges from the robot's physical embodiment, a unique style may also arise from the LLM's characteristics, influenced by our system design choices and the history of the interactions. \companion ~represents an attempt to unite robotics' physicality with LLMs' cognitive capabilities to develop an exploratory drawing experience.

\section{Related works}\label{related-works}
This work relates to two main areas: human-robot collaboration in the arts and machine learning for drawing.
Artists like Sougwen Chung and Liat Grayer explored live collaborations with robots to produce abstract artifacts. Chung's work focused on performance and co-performing with a robot trained on her drawing data \cite{chung_sketching_2022}, while Grayer's work investigated the deconstruction and replication of brushstrokes through a human-robot feedback loop \cite{grayver_transhuman_2019}. 
Research in robotic art has also focused on autonomous agents capable of observational drawing from life, utilizing feedback loops to mimic the behavior of a sketching artist \cite{tresset_artistically_2014}. While these systems successfully simulated the autonomy of a draftsman, they typically functioned independently of the artist once the process began. 
In contrast to these approaches, our \companion~project focuses on reintroducing the artist into the loop, shifting from observation to collaborative imagination and the exploration of figurative visual narratives through a direct dialogue between the artist and the machine.

In machine learning, several model architectures have been applied to drawing. Recurrent Neural Networks (RNN), like in sketch-RNN, have been used for sequence generation of visual primitives to complete user drawings \cite{ha_neural_2017}. Generative Adversarial Networks (GAN) and Diffusion Models, while effective for image generation, are less suitable for our goals \cite{goodfellow_generative_2020}. Their outputs often adhere to known aesthetics, and they lack the temporal, step-by-step construction process that is essential for our focus on storytelling and robot control. 
Research into human-AI co-creation has highlighted the importance of structured exploration in design spaces, as seen in systems like Luminate \cite{suh_luminate_2024}, distinguishing the iterative process of visual storytelling from the static visualization of a story.
LLMs are well-suited for this sequential drawing approach, as it can be treated as a sequential language like Scalable Vector Graphics (SVG) \cite{timofeenko_vector_2024}. While projects like LLM4SVG and AutoSketch explored this \cite{xing_empowering_2024, chin_autosketch_2025}, we found that generating SVGs often produced predictable styles. We therefore opted against this method to pursue more surprising outputs.

Most closely related to our approach are two recent projects that prioritize the physical and collaborative aspects of robotic drawing. Gomez Cubero et al. \cite{gomezcubero2021robot} shared our arts-led, practice-based methodology and the specific motivation to reintroduce the artist's hand into the robotic process. While their work focused on `technical ensembles' and process-oriented exploration through direct puppeteering and teleoperation, our \companion~extends this philosophy by replacing direct human control with an autonomous AI agent, aiming for a similar level of embodied fluidity but computationally driven.
Similarly, Bossema et al. \cite{bossema2025llmenhanced} investigated the integration of LLMs and speech interaction in robotic drawing, specifically within the context of creativity support for older adults. While their system shared our architectural combination of conversational agents and robotic arms, their generation pipeline relied on prepared SVGs for the robot to execute. In contrast, our approach leverages the LLM to directly generate vector commands and tool calls, bypassing pixel-based generation to achieve a sketch-like aesthetic that emerges directly from the collaborative dialogue rather than pre-rendered imagery.
Finally, Vinker et al. presented SketchAgent \cite{SketchAgent_Vinker2025}, a drawing system that leverages LLMs to generate sequential sketches. Given a text prompt, it produced a sequence of strokes that were rendered to a digital canvas. Instead of using SVG or tools like our system, it used a simple custom graphical language to describe Bezier curves on a low-resolution grid. The system also allowed for graphical interactions and editing via prompting. While it was not directed towards storytelling, it demonstrated a compelling direction for producing drawings with LLMs.
LLMs in robotics provide the key capability for our system. By integrating perception with language-based reasoning, LLMs enable robots to interpret human instructions and environmental context to generate action plans, which is fundamental to our interactive approach \cite{geminiroboticsteam2025geminirobotics15pushing, jeong_survey_2024,wulfmeier2023foundations,firoozi2025foundation}.
\begin{figure}
 \centering
 \includegraphics[width=1\linewidth]{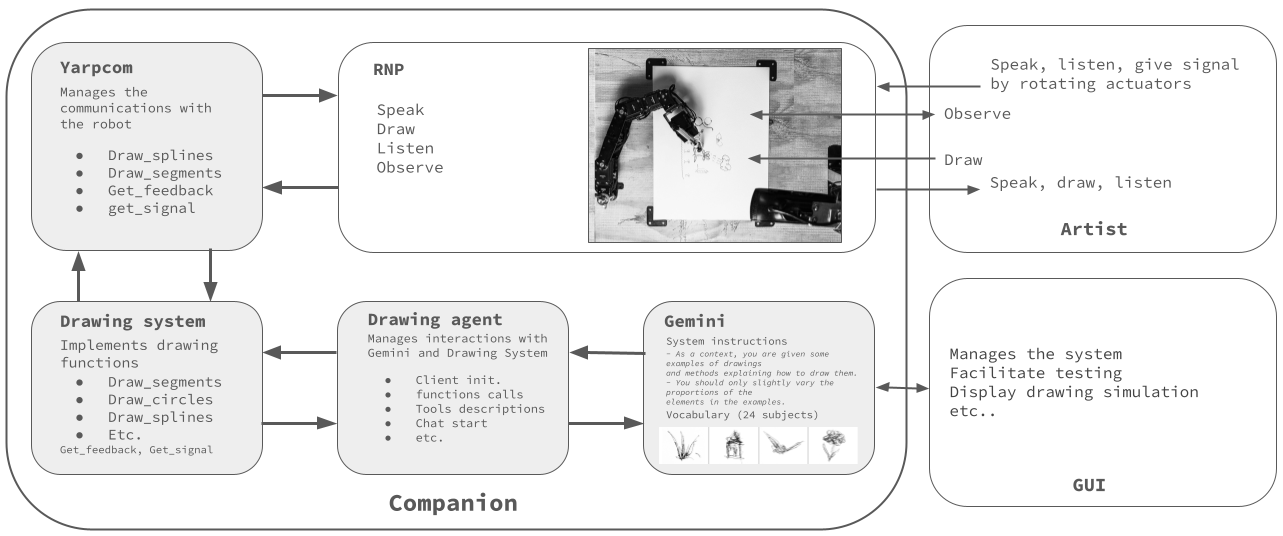}
 \caption{\companion ~overview. 
The system drives an RNP robot composed of a planar 4-jointed arm, a pan and tilt camera, a microphone, and a speaker. \companion~is composed of 3 modules: \textit{Yarpcom}, an interface with RNP, \textit{Drawing system} that implements the drawing tools such as draw line, draw segments, draw circles etc, \textit{Drawing agent }that manages the communication with Gemini via the API, including the initialization with the system instructions and context.}
 \label{fig:companion-overview}
\end{figure}

\section{Method}\label{method}
The development of our drawing partner went through an iterative process. The foundation is an embodied system with a set of drawing behaviours, allowing the agent to draw lines, paths, splines, text, circles, and other patterns. The system's capacity is described in Appendix \ref{An embodied drawing agent}. The following phases of development focused on shaping the agent's behaviour through "system instructions" design, then graphic vocabulary acquisition using In-Context Learning (ICL), and finally, the addition of voice and non-verbal communication to eliminate the need for keyboard or mouse interaction. Each of these stages is described in the following paragraphs.

The design of system instructions was used to shape the agent's behaviours. We conducted a series of experiments to craft these instructions. Different formulations of the system instructions were tested, focusing on aspects such as the agent's functions, drawing limits on the page, its understanding of the collaborative drawing task, and the inspiration for its stories. The impact of these different instructions was evaluated through artistic experimentation. An example of system instructions is presented in Appendix \ref{system instructions}.

We provided the agent with a simple graphic vocabulary to enable it to draw stories. For this purpose, we investigated the use of In-Context Learning (ICL). ICL refers to the ability of pre-trained large language models (LLMs) to perform new tasks by conditioning on demonstrations provided within the input prompt \cite{zhang_what_2023,zhu_incoro_2024}. We opted to bring a pictorial vocabulary associated with drawing methods to ICL. This approach allowed the agent to have a visual vocabulary. We experimented with different sets of drawing examples, with a variety of methods varying the details. We performed an ablation study presented in the Appendix \ref{in context learning design}.
With this working system, we first explored the system's capacities, then planned a series of experiments described in Sec. \ref{experiments}.

\subsection{Hardware Setup}
The experimental setup employs a custom-built, low-cost planar arm based on Dynamixel servos. The robot features four degrees of freedom (DoF): shoulder, elbow, wrist, and a pen-lift mechanism. It operates within a physical workspace of 170 × 130 mm.
While less precise than industrial arms or XY plotters, the robot's mechanical characteristics—specifically its compliance, backlash, and frame flexibility—are utilized as aesthetic features. By modulating movement speed, these mechanical characteristics translate into visible variations in line quality, lending the output a distinct character. 
Visual feedback is provided by a Logitech C920 camera mounted on a Dynamixel-based pan-tilt mechanism. Interaction is managed primarily through speech, recorded via the camera's microphone with silence-based auto-stop functionality. Additionally, the system supports non-verbal cues; the artist can physically rotate the robot's low-resistance joints to signal turn-taking. The low-level control is facilitated via a YARP-based framework \cite{metta_yarp_2006}, selected for its lightweight architecture compared to ROS \cite{quigley2009ros}.

\subsection{The Drawing Agent}
The core logic is managed by a Python-based agent that interfaces with the Gemini API. This module handles the system instructions, tool execution, and context management.
\subsubsection{System Instructions}
We define the agent's behavior through a structured prompt added as system instructions. These instructions delineate the agent's persona, constraints, and goals and are categorized as follows:
\begin{itemize}
\item \textbf{Embodiment and Persona:} The agent is defined as an "imaginative, story-telling drawing robot." It is instructed to possess knowledge of art history and 2D line representations, drawing inspiration from cave paintings for composition and children's drawings for narrative style.
\item \textbf{Collaborative Interaction:} The agent is explicitly instructed to facilitate a bidirectional loop. It must perceive the human's contributions on the drawing localize elements, and can request the human to draw.
\item \textbf{Stylistic Rules:} To maintain aesthetic cohesion, constraints are imposed: people are represented as stick figures, organic elements utilize splines, and slight asymmetry is preferred. To simulate line weight, the agent is instructed to, if necessary, draw thicker elements by retracing lines with a calculated offset.
\item \textbf{Geometric Constraints:} Strict physical safety margins are enforced via the prompt. The agent is aware of the simulated drawing area ( 1200 × 900 pixels) and must ensure a 40-pixel buffer between new and existing elements to prevent unwanted overlaps.
\end{itemize}
\subsubsection{In-Context Learning (ICL)}
ICL is a paradigm where a large language model performs a novel task by inferring patterns from demonstration examples provided within the input prompt at inference time, without requiring any updates to the model's parameters. For our system we constructed a library of visual examples (e.g., trees, flowers, figures) paired with procedural descriptions of how to draw them using our specific tools.
These pairs were generated via a bootstrapping method: we prompted the LLM with a reference drawing and the instruction, \textit{"Observe this drawing. Write a simple, flexible and versatile method to roughly draw it..."}. The resulting text-image pairs are uploaded to the chat session at initialization. See Appendix \ref{in context learning design}  to observe the effect of this method on the drawing style.
\subsubsection{Drawing Tools}
We utilized the Gemini API's function-calling capabilities to define a specific toolset. When the model invokes a tool, the parameters are parsed by the Drawing System and converted into paths for the robot. The tools include:
\begin{itemize}
\item \textbf{Procedural Fillers:} \texttt{draw-scribbles} and \texttt{draw-hatching} accept a polygonal boundary and a density parameter. The system fills these regions using rejection sampling (for organic scribbles) or geometric intersection (for cross-hatching).
\item \textbf{Primitives:} \texttt{draw-circles}, \texttt{draw-segments}, and \texttt{draw-text} (utilizing vector Hershey fonts) handle basic geometry.
\item \textbf{Organic Curves:} \texttt{draw-splines} and \texttt{draw-path} accept lists of keypoints, which are interpolated using cubic splines to produce fluid motions.
\item \textbf{Composite Tools:} \texttt{draw-scribbly-splines} generates a path defined by a randomized "scribble" oscillating along a central spline trajectory, mimicking a sketching style.
\end{itemize}
\subsection{Perception}
The system employs a simple perception pipeline. When the user has added lines the camera captures the drawing. This image undergoes perspective rectification using OpenCV's homography module. Contrast Limited Adaptive Histogram Equalization (CLAHE) is applied to ensure faint pen marks made by the human are visible to the model. The system also maintains a simulated version of the drawing that is used for feedback when there is no artist intervention. 
\subsection{Interaction Loop}
The interaction follows a multi-turn structure:
\begin{enumerate}
\item \textbf{Input:} When the arm's shoulder is rotated by 10 degrees, the system records sounds with the camera's microphone. Upon silence detection, the audio is saved. If the user has physically interacted with the drawing, they rotate the camera to trigger the camera frame capture, the image is rectified, and uploaded alongside the audio recording. Otherwise the simulated drawing is used.
\item \textbf{Reasoning:} The LLM processes the user's audio recording alongside the visual state (camera image or simulated drawing).
\item \textbf{Action:} The model outputs a combination of conversational text and tool calls.
\item \textbf{Execution:} Text is synthesized into speech using a text-to-speech engine (gTTS or eSpeak) configured with a non-English language parameter (e.g., French or Spanish) to read the English text. This technical mismatch generates a distinct 'non-native' accent, contributing to the agent's unique persona and differentiating it from standard service-oriented AI assistants (see Section \ref{embodiement_companionship}). The tool calls trigger both the drawing on the simulated version and the robots' actions.
\end{enumerate}

\section{Experiments}\label{experiments}
As this project originates from a professional art studio rather than a strictly academic engineering context, our evaluation methodology prioritizes "real-world" experimentation over procedural quantitative metrics. We opted for a practice-based and expert-based approach designed to assess the system’s capacity for visual storytelling, aesthetic variation, and embodied collaboration.

We first performed an ablation study to isolate the contribution of ICL to the style of the drawings (Experiment \ref{in context learning design}). To assess \companion's abilities we explored the system's abilities in three distinct roles: as a directed assistant following a human "artistic director" (Experiment \ref{artistic-director}); as an autonomous illustrator interpreting fables and abstract texts (Experiment \ref{illustrator}); and as a partner engaged in an interactive, shared drawing session (Experiment \ref{two-hands}). Furthermore, we observe the impact of the underlying technology by evaluating the evolution of drawing quality across different LLM versions (Experiment \ref{evolution-models-exp}). Finally, to address the subjectivity inherent in artistic production, we validate our results quantitatively through the Consensual Assessment Technique (CAT) utilizing a panel of domain experts (Sec. \ref{sec:evaluation}).

The experiments presented in the appendices were used to guide the system's development. The series presented below is designed to explore and assess the system's capacities. 

\subsection{A Human Artistic Director}\label{artistic-director}
\begin{figure}
 \centering
 \includegraphics[width=.7\linewidth]{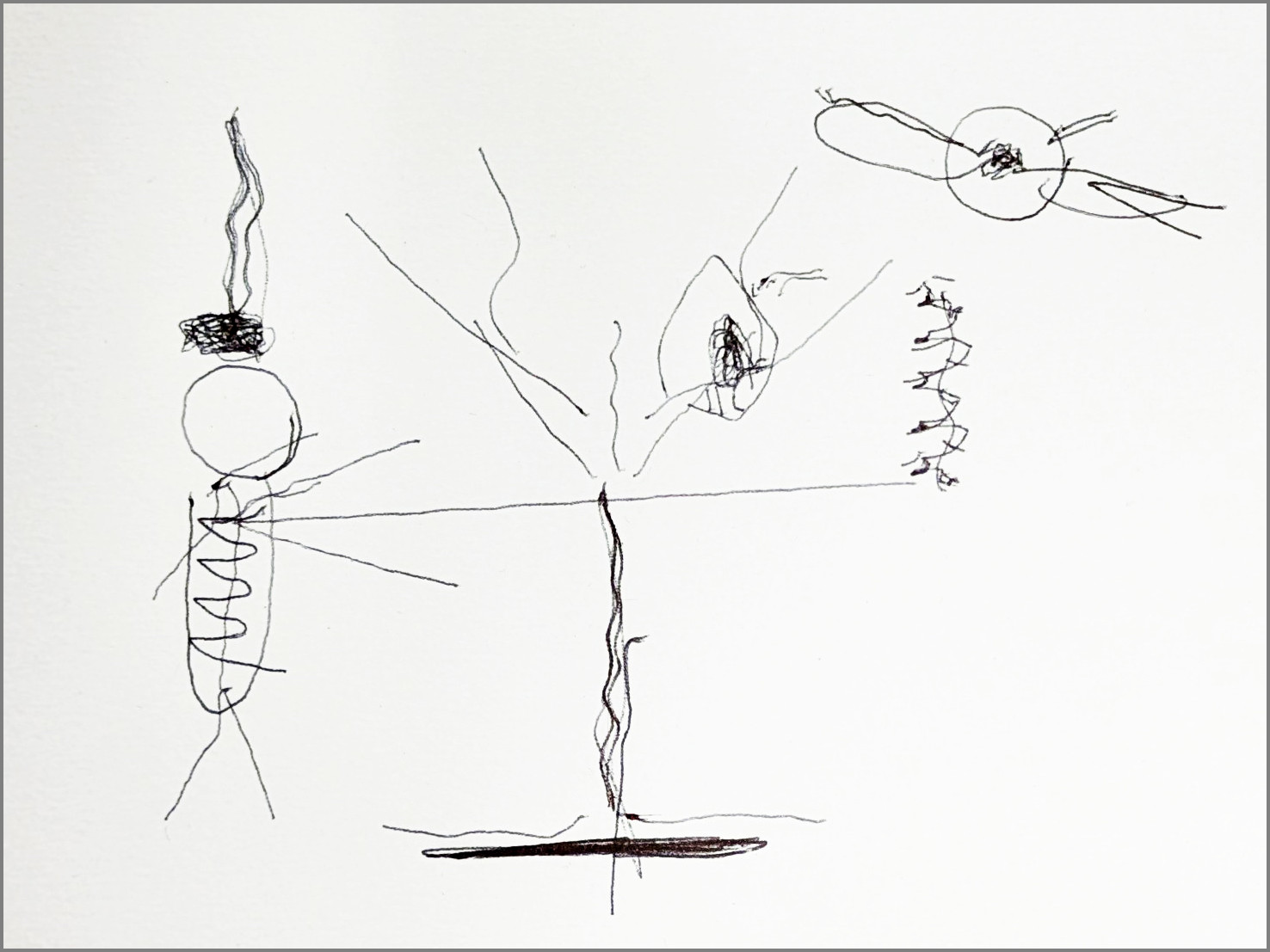}
 \caption{The feather illustrates the 'human artistic director' scenario. Note the elongated arm drawn by the robot to 'catch' the feather, deviating from the user's suggestion to reposition the character.  (Experiment \ref{artistic-director})}
 \end{figure}
 \label{fig:the-feather}
 
In this scenario, the artist did not intervene in the drawing or story; they work as an artistic adviser. The story \companion ~imagined was as follows: a character sees a bird in a tree and points at it. The bird becomes scared and flees, losing a magical feather. The character catches it and places it on their hat.
During the process, we gave some directions, such as adding clothes and a hat to the character or adding more details to the feather. These were followed. However, some ideas were politely rejected, like using dashed lines and arrows to depict time, or ignored, such as when we suggested drawing the character in another position to catch the feather instead; \companion ~drew a very long arm.

\subsection{The system as an illustrator}\label{illustrator}
During this experiment, we explored \companion's capacity to illustrate well-known stories such as "The Hare and the Turtle", "The Crow and the Fox", the sentence "Try, fail, try again, fail better." by Samuel Beckett, and a song "Tous les garçons et les filles" by Françoise Hardy. In this scenario, we did not give any indication, only a description of the task. After that, we asked: "\textit{What will you draw next?}" until the system considered the illustration completed. 
\begin{figure}
 \centering
 \includegraphics[width=0.7\linewidth]{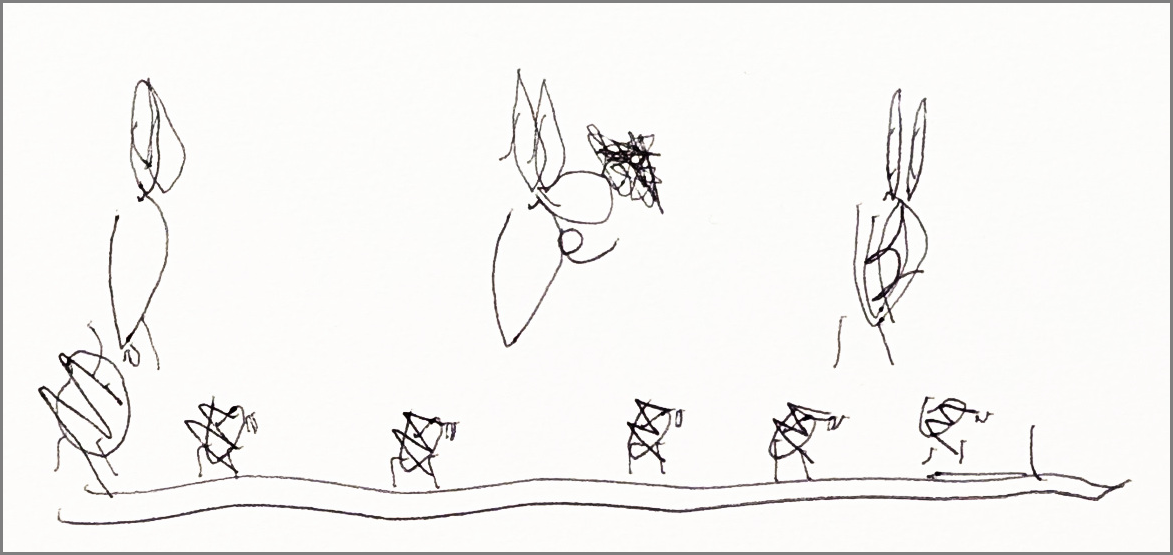}
\caption{The Hare and the Turtle, (Experiment \ref{hare-turtle-exp}). A sequential representation of the fable. The turtle (multiple figures) progresses steadily across the page, while the hare stops early. The vertical line on the right represents the finish line}
\end{figure}\label{fig:hare-turtle}
\subsubsection{The Hare and Turtle}\label{hare-turtle-exp}
In this case, even though there is no turtle or hare in the ICL vocabulary, the result is recognisable. The unfolding of time goes from left to right. The turtle is repeated six times, and the hare only three times, showing that the turtle does not stop, but the hare does. The vertical line represents the finishing line. The last hare is scribbled over to show unhappiness (see Fig. \ref{fig:hare-turtle}).
\begin{figure}
 \centering
 \includegraphics[width=0.9\linewidth]{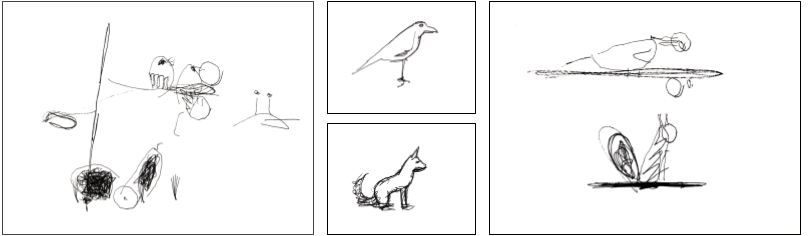}
 \caption{The Crow and the Fox (Experiment \ref{crow-fox}). Left: The first attempt. Middle: The two drawings shown by the user. Right: the agent's second attempt, with improved crow and fox depiction.}
 \label{fig:crow-fox}
\end{figure}
\subsubsection{The Crow and the Fox}\label{crow-fox}
To illustrate this story, \companion ~encountered difficulties drawing the fox and the crow. We experimented with different strategies to improve the situation. The most successful was to show the drawing of a fox and ask to copy it, then the same for the crow. During the copying process, we gave feedback on how to improve the drawing. We then indicated that we removed the page and that it should illustrate Jean de la Fontaine's tale. The result was enhanced (Fig. \ref{fig:crow-fox}). The story unfolds visually with the cheese drawn three times, in the crow's beak, falling, and in the fox's mouth.

\begin{figure}
 \centering
 \includegraphics[width=0.8\linewidth]{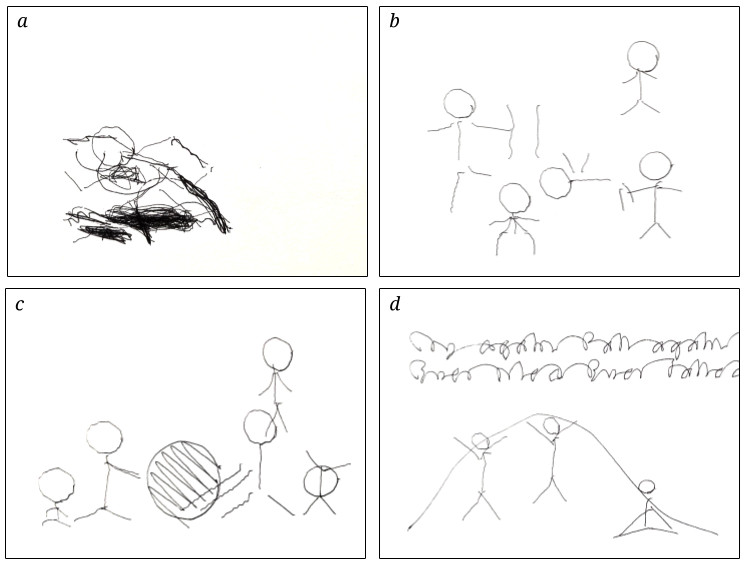}
 \caption{ Try again. Fail again. Fail better (Experiment \ref{try-fail}), the four images show the different ways \companion illustrated the sentence}
 \label{fig:try-again}
\end{figure}
\subsubsection{Try, Fail, Try Again, Fail Better}\label{try-fail}
We asked \companion ~to illustrate Samuel Beckett's well-known sentences from "Worstward Ho": "Ever tried? Ever failed? No matter. Try again. Fail again. Fail better.". We asked four times to evaluate the various interpretations it would give us. Contrary to the tales here, there are no defined subjects (Fig. \ref{fig:try-again}). In drawing \textit{a)}, the different stages are overlaid. In drawing \textit{b)}, multiple stick figures in various states are represented without a specific order, perhaps to depict the endless repetition. In \textit{c)} An accessory, a large boulder, is used, with the figures in different positions. In \textit{d)} the context, the hill used in the landscape seems to represent success and failure. 
\begin{figure}
 \centering
 \includegraphics[width=0.7\linewidth]{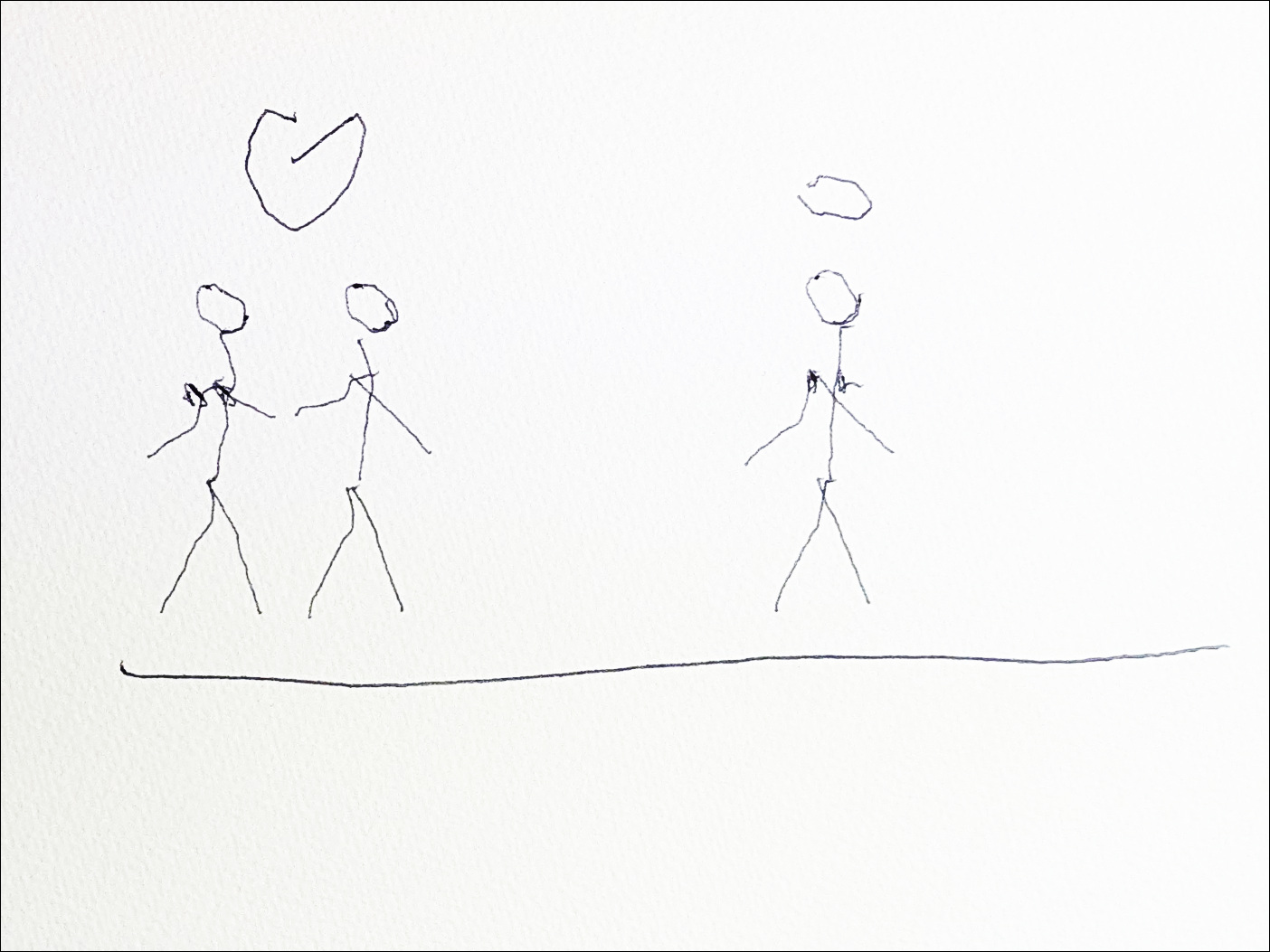}
 \caption{Tous les garçons et les filles (Experiment \ref{tous-les-garcons}). An illustration of Françoise Hardy’s song lyrics. The drawing features figures holding hands and hearts, drawn sequentially until the system deemed the drawing complete.}
 \label{fig:touslesgf}
\end{figure}
\subsubsection{Tous les garçons et les filles}\label{tous-les-garcons}
For this experiment, we played the beginning of Françoise Hardy's song and asked \companion ~to illustrate it. 
We then asked to continue until it considers the drawing complete (Fig. \ref{fig:touslesgf}).
\subsection{A story with two hands}\label{two-hands}
\begin{figure}
 \centering
 \includegraphics[width=1\linewidth]{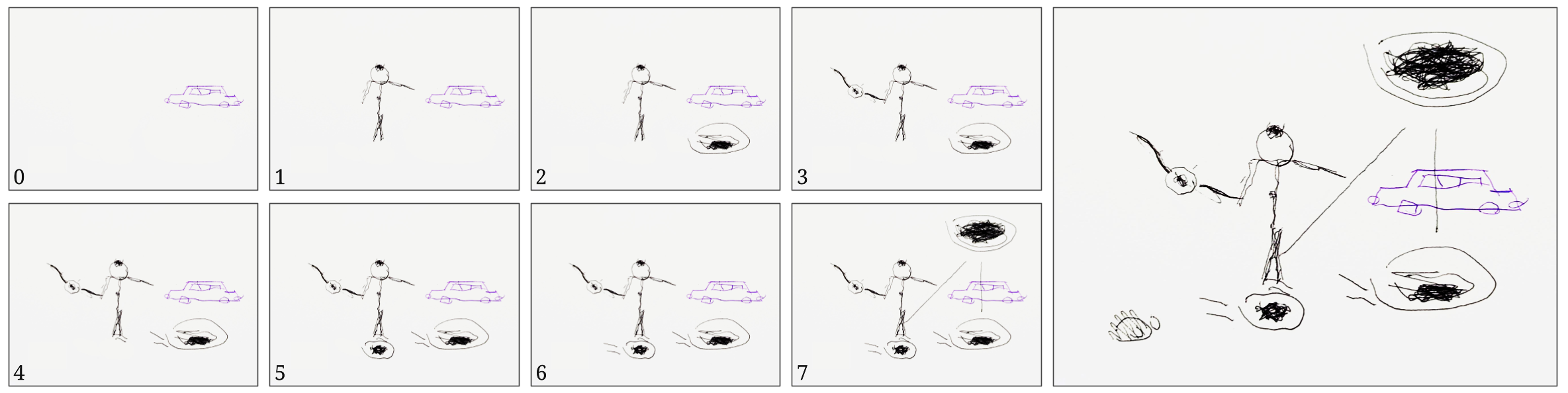}
 \caption{ The Floating Car (Experiment \ref{two-hands}). A co-produced drawing where the artist drew a car high on the page. The Companion interpreted this as a 'flying car' and added a stick figure floating upward on a cloud to match the narrative."}
 \label{fig:floating-car}
\end{figure}
\subsubsection{The artist initiated the drawing, and \companion ~imagined the story}
In this case, the artist began by drawing a car. As the vehicle was drawn high on the page, \companion ~interpreted it as floating.
\companion's description:
\textit{On the lower left, a patch of sketchy grass signifies the ordinary ground left behind. From this point, the story takes flight! A stick figure, dressed in hatched clothes and potentially partnered with a startled bird (or perhaps just pointing in astonishment), is floating upwards on a personal, scribbly cloud. Following a similar trajectory, the simple car you first drew also ascends on its own cloud, moving towards the upper right. Both the figure and the car are being drawn towards a large, dark, mysterious cloud formation that dominates the top right corner. Faint trails indicate their path into this cloud, suggesting that they have vanished into its depths, continuing their journey into the unknown. The scene captures a moment of unexpected magic, departure, and whimsical mystery.}
The title as Haiku:
\textit{Car floats up so high
Figure follows to the cloud
Where do they both go?}

\begin{figure}
 \centering
 \includegraphics[width=0.9\linewidth]{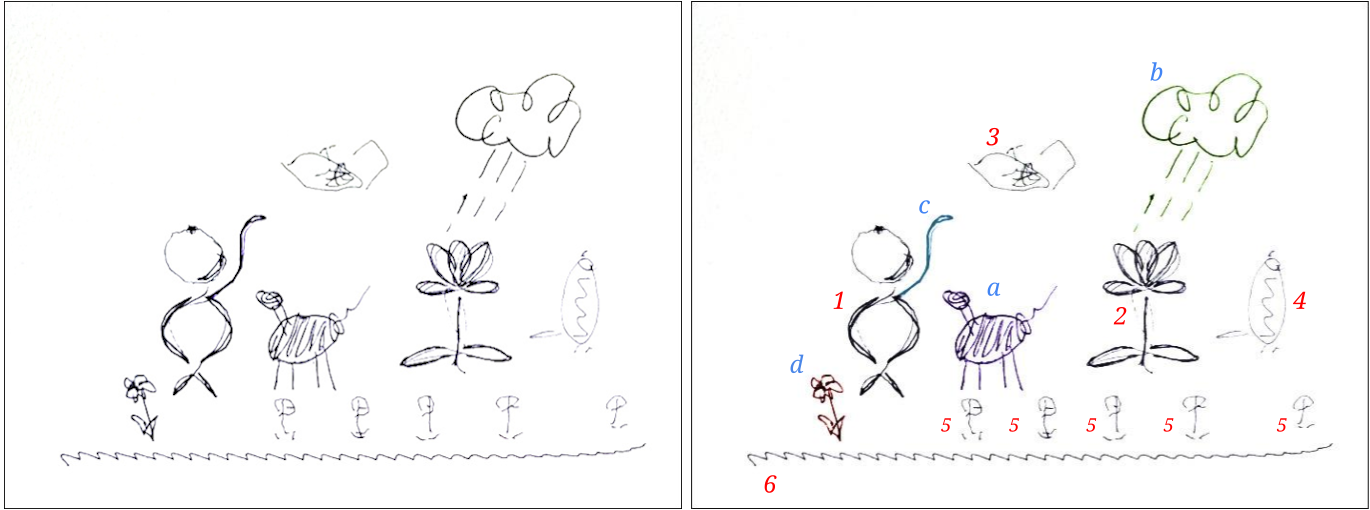}
 \caption{The sheep, the bird and the six flowers (Experiment \ref{cooperative-exp}). 
 Collaborative drawing showing a sheep, bird, and flowers. Elements 1-6 were drawn by \companion, and elements a-d were added by the artist.}
 \label{fig:sheep-flowers}
\end{figure}
\subsubsection{The artist and \companion ~work on the drawing and story together}\label{cooperative-exp}
In this scenario, the artist and \companion ~cooperated on a single drawing (Fig. \ref{fig:sheep-flowers}). The \companion ~began by drawing a character (1), to which the artist adds an animal (a). The \companion ~then drew a flower (2), followed by the artist drawing a cloud with rain (b). Next, the \companion's drawing of a flying bird (3) led the artist to add the character's arm pointing towards it (c). The \companion ~proceeded to draw the bird landed (4). Subsequently, the artist drew a flower (d) and asked the \companion ~to draw five additional ones. Finally, the collaboration involved the artist requesting, and the \companion ~adding, hatching at the bottom of the drawing (6). 
\begin{figure}
 \centering
 \includegraphics[width=0.9\linewidth]{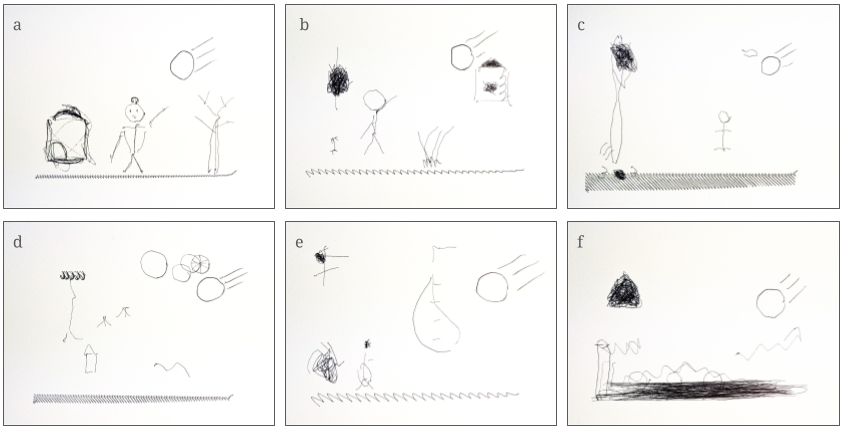}
 
 \caption{Comparison of \companion's output for the same task (interpreting an artist-drawn 'asteroid', drawing an imagined story, and adding ground hatching) across different Gemini model versions, illustrating the evolution of drawing capability. Models shown are: a) Gemini-2.5-pro-preview, b) Gemini-2.0-pro, c) Gemini-2.0-flash, d) Gemini-2.0-flash-lite, e) Gemini-1.5-pro, f) Gemini-1.5-flash (Experiment \ref{evolution-models-exp})}
 \label{fig:evolution-models}
\end{figure}
\subsection{The evolution across models}\label{evolution-models-exp}
To assess how language model evolution impacts \companion's drawing, we conducted a consistent experiment across different Gemini models, shown in Figure \ref{fig:evolution-models}. The task was: the artist draws an element (a circle with three trailing lines, suggesting an asteroid). \companion ~observes this mark, imagines a story based on it, and performs the drawing. Lastly, \companion ~is asked to add a requested element, like ground hatching. In this experiment, we can observe a clear evolution from Gemini-1.5-flash to Gemini-2.5-pro. The earliest model producing more abstract drawings, with the subject being unrecognisable to being clear with the latest system. With the earliest model \companion ~confidently states that it is drawing each subject. This could either presents an interesting case of so-called "Hallucination", or just that the system doesn't have the capacity to draw the subject.

\begin{figure}
 \centering
 \includegraphics[width=0.9\linewidth]{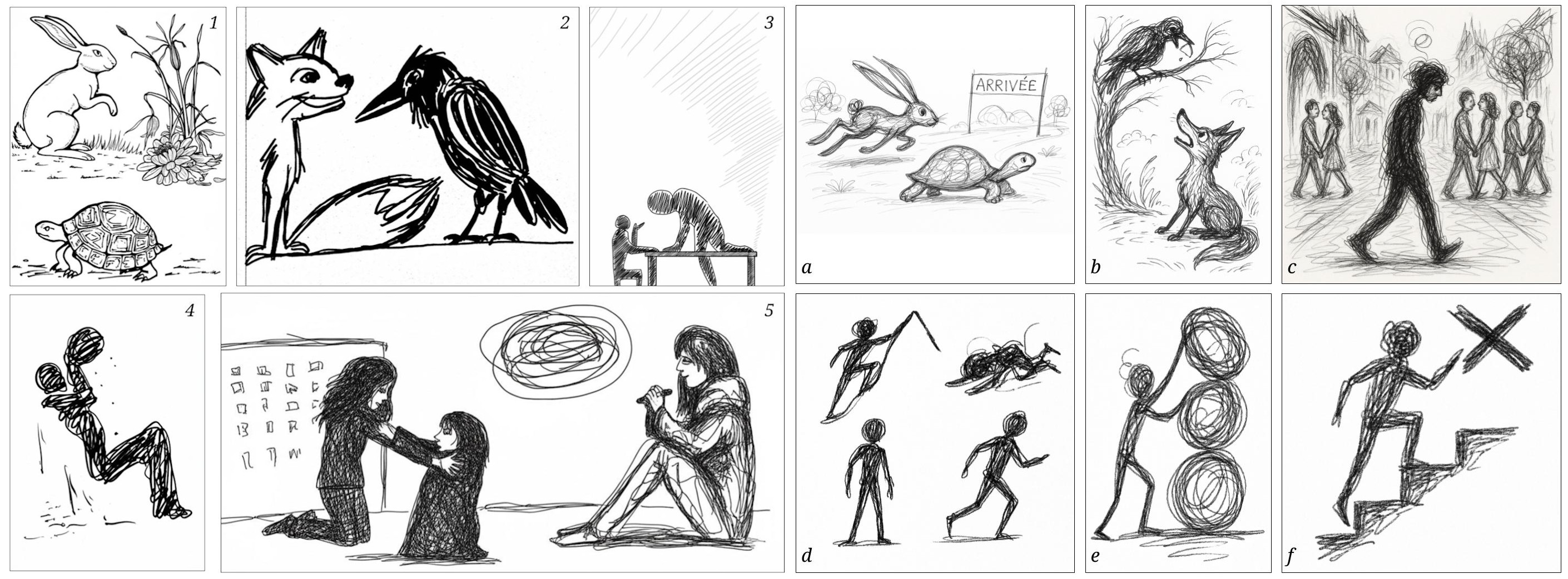}
 \caption{LLM-diffusion models as illustrators
 \textit{1-5} - Gemini-2.0-flash-exp
 \textit{a-f} - ChatGPT 4o (Experiment \ref{diffusion-llm})}
 \label{fig:diffusion-illustrator}
\end{figure}

\subsection{Comparison with text to image models}\label{diffusion-llm}
We conducted the Illustrator experiments with Gemini-2.0-flash-exp and ChatGPT-4o (Fig. \ref{fig:diffusion-illustrator}). We added the scribbly vocabulary at the beginning of the chat as examples and then asked the system to illustrate the fables, sentences and song in one image. For the two fables, we tried different strategies to depict the full story in one photo for the systems, but without success. Both systems produced illustrations for the song and Beckett's sentences. The influence of the vocabulary's style is apparent with both models, but the aesthetic and stylisation remain traditional.

\begin{figure}
  \centering
  \includegraphics[width=1\linewidth]{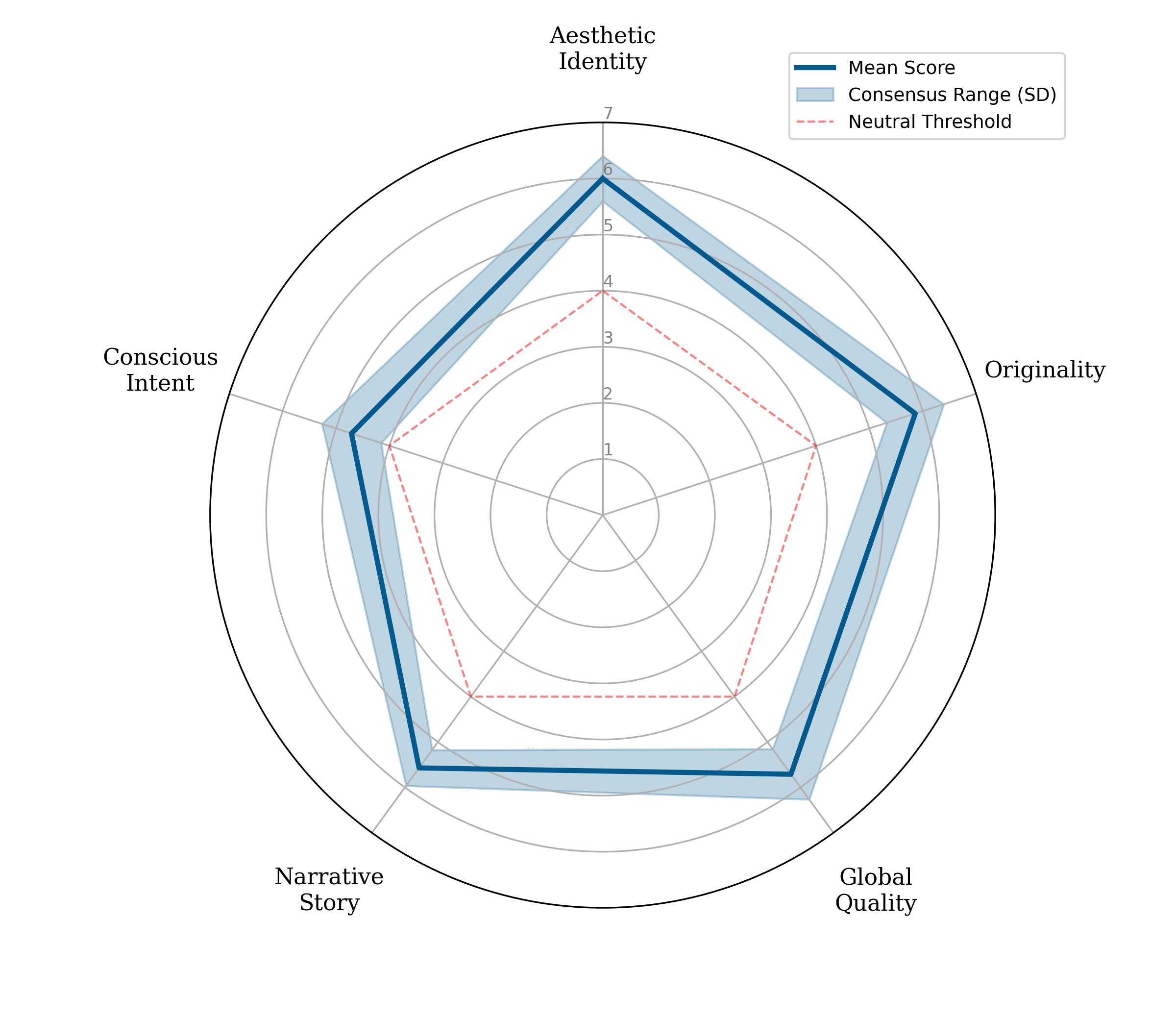}
  \caption{Expert evaluation profile (N=7) across five dimensions. The solid blue line represents the Mean score (1=Low, 7= High). The shaded region represents the Standard Deviation, indicating the level of consensus among the judges. The system shows a distinct profile with high agreement on Aesthetic Identity (narrow band) and slightly more variation in Conscious Intent (wider band).}
  \label{fig:cat_radar}
\end{figure}
\subsection{Evaluation and Expert Consensus}\label{sec:evaluation}
To address the subjectivity inherent in evaluating artistic production and to validate the system’s capabilities beyond the authors' perspective, we employed the Consensual Assessment Technique (CAT)~ \cite{amabile1982social}. We assembled a panel of seven domain experts, comprising five professional curators, a collector, and an artist who uses robotics. All participants were familiar with the first author's prior works; this familiarity ensured that their evaluation focused on the specific contributions of the \companion~system rather than the novelty of robotic drawing.
The panel was provided access to a high-resolution portfolio of the system's output and video documentation of the embodied interaction. They were instructed to evaluate the work as a cohesive body of art using a 1--7 Likert scale across five dimensions. We calculated the Mean ($M$), Standard Deviation ($SD$), and the Within-Group Interrater Reliability ($r_{wg}$) to measure consensus.
\subsubsection{Results}
The results are visualized in Figure~\ref{fig:cat_radar}. The system achieved its highest rating in Aesthetic Identity ($M=6.0, SD=0.81$). The panel reached moderate-to-strong consensus ($r_{wg}=0.67$) that the drawings exhibited a distinct, cohesive style. Qualitative feedback compared the aesthetic to ``Pre-historic petroglyphs,'' ``Paul Klee,'' and ``Basquiat,'' describing the style as ``primitive,'' ``bold,'' and possessing a ``visual language'' unique to the machine.

\textbf{Global Quality} ($M=5.71$) and \textbf{Originality} ($M=5.86$) were also rated highly, validating the system’s output as suitable for professional exhibition. One expert noted that the ``hesitant, accumulated lines'' created a ``strong narrative layer,'' transforming the robot from a plotter into an observer ``learning to see.''

\textbf{Intent and Agency:} The metric for \textbf{Conscious Presence/Intent} received the lowest score ($M=4.71, SD=1.11$). This lower score reflects the experts' professional discernment regarding machine agency. However, the qualitative comments highlighted a shift in perception: one curator observed that ``authorship feels shared, not simply delegated,'' noting that the interaction created an ``intimate relationship'' between the artist and the agent.

\section{Discussion}\label{discussion}
This paper introduces the '\companion', an embodied AI agent designed for collaborative visual storytelling, motivated by a desire to reintroduce the artist's direct presence into the robotic drawing process and explore unexpected artistic territories, both aesthetically and narratively. Our experiments demonstrate the feasibility of using LLMs combined with In-Context Learning (ICL), function calling, and embodied interaction (speech, physical drawing) to create a partner capable of engaging in explorative drawing.

\subsection{A Speech-Driven Conversational Drawing Agent}
The core system we have developed is a speech-driven, conversational embodied drawing agent that can be used by indicating what to draw using its implemented tools. 

\subsection{Defining a Consistent Style-Space}
Our ablation study on ICL (Appendix \ref{in context learning design}) revealed a distinction between visual recognition and procedural execution. The failure of the 'Visual Context Only' condition—resulting in fragmented or abstract outputs—suggests that while LLMs possess internal visual knowledge, they struggle to map pixel data directly to the specific constraints of our drawing tools. There remains a 'translation gap' between seeing an image and understanding the sequence of motor primitives required to construct it.
However, the success of the 'Full Context' condition (pairing sketches with methods) indicates that the model seems capable of 'Procedural Style Transfer.' By observing the methods, the agent did not memorize shapes but inferred implicit stylistic rules—such as the tendency to anchor figures with hatched ground lines or use specific curves for organic forms. The agent successfully applied these stylistic grammars to unseen subjects, demonstrating a generalization capability. This suggests that for embodied creative agents, "style" is best communicated not just as a visual target, but as a logic of action.

\subsection{Collaborative Partnership and Emergent Agency}
The goal is to shift from detached control to playful involvement, seeking surprise and new directions. The experiments suggest working with \companion~leads to unexpected stories and depictions. In scenarios where the artist worked as a director (Sec. \ref{artistic-director}) or when collaborating directly on a shared drawing (Sec. \ref{two-hands} ), the \companion~exhibited a degree of agency. It interpreted verbal prompts, integrated visual vocabulary (influenced by ICL), and contributed narrative elements, sometimes following instructions, other times offering unexpected interpretations, like the "floating car" (Fig. \ref{fig:floating-car}) or deviating from suggestions, like drawing the long arm instead of repositioning the character (Sec. \ref{artistic-director}). This unpredictability, while sometimes requiring negotiation (e.g., the copying needed for the Crow/Fox in Sec. \ref{crow-fox}), aligns with our aim for the system to pull the artist into unforeseen territories, moving beyond habitual patterns. The successful turn-taking in the collaborative drawing (Fig. \ref{fig:sheep-flowers}) highlights its potential as a partner, responding to and building upon the artist's contributions.

\subsection{Visual Storytelling and Narrative Representation}
The \companion~demonstrated competence in illustrating narratives, from fables (Sec. \ref{hare-turtle-exp},\ref{crow-fox} ) to song lyrics (Sec. \ref{tous-les-garcons}). A key finding is the variety of strategies the \companion~used to depict the passage of time and action. These included:
Sequential Representation: Repeating characters across space (the turtle's steady progress, Fig \ref{fig:hare-turtle}).
Positional Cues: Showing elements in different states or locations within the narrative sequence, the cheese moving from beak to mouth (Fig. \ref{fig:crow-fox}).
Figurative Representation: Adding or modifying elements that function symbolically, the elongated arm suggesting action without repositioning (Fig. \ref{fig:the-feather}).
These emergent strategies suggest the LLM, guided by system instructions and ICL, can access and translate narrative concepts into sequences of drawing actions, offering diverse visual outputs.
The observed progression across Gemini model versions (Sec. \ref{evolution-models-exp}, Fig. \ref{fig:evolution-models}) indicates that the underlying LLM's capabilities impact the legibility and complexity of the final output, even if the narrative intents are comparable.

\subsection{Embodiment and Companionship}\label{embodiement_companionship}
Etymologically, a companion is someone we share bread with. They are present with us, someone with whom we can communicate, trust, work, and travel, and with whom we share empathy, respect, reciprocity, authenticity, and common goals.
When working with a non-embodied LLM, it is clear that the agent is cloud-based and remote. With a robot, the system is present in the same physical and temporal space as us even if parts of its "intelligence" remains remote. In our context, we also share the workspace; we are building something together, we have a shared aim. The use of vocal communication also removes a distance and an interface: writing.
In \companion, we have opted to add an accent to the speech. While there is now a large variety of voices in text-to-speech models, adding an accent gives further individuality to the agent. Without it, the voice gives the impression of conversing with a professional, someone with a social distance.
To enhance the agent's perceived autonomy, we eliminated the need for keyboard or mouse interaction by communicating through physical actions, such as touching the arm or the camera's motor.

\subsection{Validity and Cultural Impact}
A critical finding from our evaluation is the validation of the system's "Aesthetic Identity" ($M=6.0/7$). Given that the expert panel was already familiar with the first author's previous body of work, this high score indicates that \companion~has successfully generated a new, emergent style distinct from the artist’s previous drawing robots [Anonymized]. The experts identified a "child-like" or "naïve" quality in the mark-making—comparisons to "children's drawings" or "early sketches" were frequent. Far from being a critique, the panel identified this as the system's strength. As one expert noted, this "cryptic" quality allows the audience to "project their own version of the story. 
The system’s production has achieved professional recognition. A small but reputable art publisher will release a book with 24 drawings produced with \companion , the same drawings and two participatory installations based on the technology we developed will be part of a solo show in an art center in 2026. These commitments serve as external validation of the system's capacity to produce artifacts that hold cultural and aesthetic value.
\section{Conclusion}\label{conclusion}
This paper introduced '\companion', an embodied AI agent developed for collaborative visual storytelling. Motivated by a desire to reintroduce the artist's presence and explore unpredictable aesthetic territories, we integrated a physical drawing robot with LLMs, leveraging In-Context Learning with a visual vocabulary and intuitive interaction methods. Our experiments demonstrated the system's feasibility as a cooperative partner, capable of interpreting prompts, contributing narrative elements with emergent agency, and generating physical drawings with a unique style distinct from purely digital outputs. This agent represents a step towards using embodied AI not just as a tool, but as a partner for shared, explorative visual production on paper. Apart from the creation of a series of artworks based on the system and its output, future work will investigate the possibility of using the system for observational drawing, and how to get the system to develop its tools, and to implement a system that enables individuals with arm-hand motor disabilities to draw. 
\section*{Acknowledgements}
This work has been partly funded by the EACVA (Embodied Agents in Contemporary Visual Art: \href{https://eacva.org/}{eacva.org}) project jointly led by Goldsmiths --- under grant no. AH/X002241/1 from the UKRI/AHRC ---  and the University of Konstanz --- under grant no. 508324734 from the Deutsche Forschungsgemeinschaft (DFG).
\printbibliography

@misc{geminiroboticsteam2025geminirobotics15pushing,
      title={Gemini Robotics 1.5: Pushing the Frontier of Generalist Robots with Advanced Embodied Reasoning, Thinking, and Motion Transfer}, 
      author={Gemini Robotics Team and Abbas Abdolmaleki and Saminda Abeyruwan and Joshua Ainslie and Jean-Baptiste Alayrac and Montserrat Gonzalez Arenas and Ashwin Balakrishna and Nathan Batchelor and Alex Bewley and Jeff Bingham and Michael Bloesch and Konstantinos Bousmalis and Philemon Brakel and Anthony Brohan and Thomas Buschmann and Arunkumar Byravan and Serkan Cabi and Ken Caluwaerts and Federico Casarini and Christine Chan and Oscar Chang and London Chappellet-Volpini and Jose Enrique Chen and Xi Chen and Hao-Tien Lewis Chiang and Krzysztof Choromanski and Adrian Collister and David B. D'Ambrosio and Sudeep Dasari and Todor Davchev and Meet Kirankumar Dave and Coline Devin and Norman Di Palo and Tianli Ding and Carl Doersch and Adil Dostmohamed and Yilun Du and Debidatta Dwibedi and Sathish Thoppay Egambaram and Michael Elabd and Tom Erez and Xiaolin Fang and Claudio Fantacci and Cody Fong and Erik Frey and Chuyuan Fu and Ruiqi Gao and Marissa Giustina and Keerthana Gopalakrishnan and Laura Graesser and Oliver Groth and Agrim Gupta and Roland Hafner and Steven Hansen and Leonard Hasenclever and Sam Haves and Nicolas Heess and Brandon Hernaez and Alex Hofer and Jasmine Hsu and Lu Huang and Sandy H. Huang and Atil Iscen and Mithun George Jacob and Deepali Jain and Sally Jesmonth and Abhishek Jindal and Ryan Julian and Dmitry Kalashnikov and M. Emre Karagozler and Stefani Karp and Matija Kecman and J. Chase Kew and Donnie Kim and Frank Kim and Junkyung Kim and Thomas Kipf and Sean Kirmani and Ksenia Konyushkova and Li Yang Ku and Yuheng Kuang and Thomas Lampe and Antoine Laurens and Tuan Anh Le and Isabel Leal and Alex X. Lee and Tsang-Wei Edward Lee and Guy Lever and Jacky Liang and Li-Heng Lin and Fangchen Liu and Shangbang Long and Caden Lu and Sharath Maddineni and Anirudha Majumdar and Kevis-Kokitsi Maninis and Andrew Marmon and Sergio Martinez and Assaf Hurwitz Michaely and Niko Milonopoulos and Joss Moore and Robert Moreno and Michael Neunert and Francesco Nori and Joy Ortiz and Kenneth Oslund and Carolina Parada and Emilio Parisotto and Amaris Paryag and Acorn Pooley and Thomas Power and Alessio Quaglino and Haroon Qureshi and Rajkumar Vasudeva Raju and Helen Ran and Dushyant Rao and Kanishka Rao and Isaac Reid and David Rendleman and Krista Reymann and Miguel Rivas and Francesco Romano and Yulia Rubanova and Peter Pastor Sampedro and Pannag R Sanketi and Dhruv Shah and Mohit Sharma and Kathryn Shea and Mohit Shridhar and Charles Shu and Vikas Sindhwani and Sumeet Singh and Radu Soricut and Rachel Sterneck and Ian Storz and Razvan Surdulescu and Jie Tan and Jonathan Tompson and Saran Tunyasuvunakool and Jake Varley and Grace Vesom and Giulia Vezzani and Maria Bauza Villalonga and Oriol Vinyals and René Wagner and Ayzaan Wahid and Stefan Welker and Paul Wohlhart and Chengda Wu and Markus Wulfmeier and Fei Xia and Ted Xiao and Annie Xie and Jinyu Xie and Peng Xu and Sichun Xu and Ying Xu and Zhuo Xu and Jimmy Yan and Sherry Yang and Skye Yang and Yuxiang Yang and Hiu Hong Yu and Wenhao Yu and Wentao Yuan and Yuan Yuan and Jingwei Zhang and Tingnan Zhang and Zhiyuan Zhang and Allan Zhou and Guangyao Zhou and Yuxiang Zhou},
      year={2025},
      eprint={2510.03342},
      archivePrefix={arXiv},
      primaryClass={cs.RO},
      url={https://arxiv.org/abs/2510.03342}, 
}

@article{wulfmeier2023foundations,
  title={Foundations for transfer in reinforcement learning: A taxonomy of knowledge modalities},
  author={Wulfmeier, Markus and Byravan, Arunkumar and Bechtle, Sarah and Hausman, Karol and Heess, Nicolas},
  journal={arXiv preprint arXiv:2312.01939},
  year={2023}
}

@article{firoozi2025foundation,
  title={Foundation models in robotics: Applications, challenges, and the future},
  author={Firoozi, Roya and Tucker, Johnathan and Tian, Stephen and Majumdar, Anirudha and Sun, Jiankai and Liu, Weiyu and Zhu, Yuke and Song, Shuran and Kapoor, Ashish and Hausman, Karol and others},
  journal={The International Journal of Robotics Research},
  volume={44},
  number={5},
  pages={701--739},
  year={2025},
  publisher={SAGE Publications Sage UK: London, England}
}

@inproceedings{quigley2009ros,
  title={ROS: an open-source Robot Operating System},
  author={Quigley, Morgan and Conley, Ken and Gerkey, Brian and Faust, Josh and Foote, Tully and Leibs, Jeremy and Wheeler, Rob and Ng, Andrew Y and others},
  booktitle={ICRA workshop on open source software},
  volume={3},
  number={3.2},
  pages={5},
  year={2009},
  organization={Kobe}
}

@article{amabile1982social,
  title={Social Psychology of Creativity: A Consensual Assessment Technique},
  author={Amabile, Teresa M.},
  journal={Journal of Personality and Social Psychology},
  volume={43},
  number={5},
  pages={997--1013},
  year={1982},
  publisher={American Psychological Association}
}

@article{gomezcubero2021robot,
  title={The Robot is Present: Creative Approaches for Artistic Expression With Robots},
  author={Gomez Cubero, Carlos and Pekarik, Maros and Rizzo, Valeria and Jochum, Elizabeth},
  journal={Frontiers in Robotics and AI},
  volume={8},
  pages={662249},
  year={2021},
  publisher={Frontiers}
}

@misc{bossema2025llmenhanced,
  title={LLM-enhanced Interactions in Human-Robot Collaborative Drawing with Older Adults},
  author={Marianne Bossema and Somaya Ben Allouch and Aske Plaat and Rob Saunders},
  year={2025},
  eprint={2506.18711},
  archivePrefix={arXiv},
  primaryClass={cs.HC}
}

@InProceedings{SketchAgent_Vinker2025,
            author    = {Vinker, Yael and Shaham, Tamar Rott and Zheng, Kristine and Zhao, Alex and E Fan, Judith and Torralba, Antonio},
            title     = {SketchAgent: Language-Driven Sequential Sketch Generation},
            booktitle = {Proceedings of the Computer Vision and Pattern Recognition Conference (CVPR)},
            month     = {06},
            year      = {2025},
            pages     = {23355-23368}
        }

@inproceedings{suh_luminate_2024,
  title={Luminate: Structured Generation and Exploration of Design Space with Large Language Models for Human-AI Co-Creation},
  author={Suh, Sangho and Min, Bryan and Sunkara, Sajjadur and Nasihati, Arman and Li, Toby Jia-Jun},
  booktitle={Proceedings of the CHI Conference on Human Factors in Computing Systems},
  year={2024}
}

@incollection{chung_sketching_2022,
	title = {Sketching symbiosis: {Towards} the development of relational systems},
	booktitle = {The {Language} of {Creative} {AI}: {Practices}, {Aesthetics} and {Structures}},
	publisher = {Springer},
	author = {Chung, Sougwen},
	year = {2022},
	pages = {259--276},
}

@article{moura_machines_2016,
	title = {Machines that make art},
	journal = {Robots and Art: Exploring an Unlikely Symbiosis},
	author = {Moura, Leonel},
	year = {2016},
	pages = {255--269},
}

@incollection{grba_transparency_2023,
	title = {The transparency of reason: ethical issues of {AI} art},
	booktitle = {Handbook of critical studies of artificial intelligence},
	publisher = {Edward Elgar Publishing},
	author = {Grba, Dejan},
	year = {2023},
	pages = {504--514},
}

@article{team_gemini_2025,
	title = {Gemini: {A} {Family} of {Highly} {Capable} {Multimodal} {Models}},
	shorttitle = {Gemini},
	url = {http://arxiv.org/abs/2312.11805},
	doi = {10.48550/arXiv.2312.11805},
	urldate = {2025-05-28},
	author = {Team, Gemini and Anil, Rohan and Borgeaud, Sebastian and Alayrac, Jean-Baptiste and Yu, Jiahui and Soricut, Radu and Schalkwyk, Johan and Dai, Andrew M. and Hauth, Anja and Millican, Katie and Silver, David and Johnson, Melvin and Antonoglou, Ioannis and Schrittwieser, Julian and Glaese, Amelia and Chen, Jilin and Pitler, Emily and Lillicrap, Timothy and Lazaridou, Angeliki and Firat, Orhan and Molloy, James and Isard, Michael and Barham, Paul R. and Hennigan, Tom and Lee, Benjamin and Viola, Fabio and Reynolds, Malcolm and Xu, Yuanzhong and Doherty, Ryan and Collins, Eli and Meyer, Clemens and Rutherford, Eliza and Moreira, Erica and Ayoub, Kareem and Goel, Megha and Krawczyk, Jack and Du, Cosmo and Chi, Ed and Cheng, Heng-Tze and Ni, Eric and Shah, Purvi and Kane, Patrick and Chan, Betty and Faruqui, Manaal and Severyn, Aliaksei and Lin, Hanzhao and Li, YaGuang and Cheng, Yong and Ittycheriah, Abe and Mahdieh, Mahdis and Chen, Mia and Sun, Pei and Tran, Dustin and Bagri, Sumit and Lakshminarayanan, Balaji and Liu, Jeremiah and Orban, Andras and Güra, Fabian and Zhou, Hao and Song, Xinying and Boffy, Aurelien and Ganapathy, Harish and Zheng, Steven and Choe, HyunJeong and Weisz, Ágoston and Zhu, Tao and Lu, Yifeng and Gopal, Siddharth and Kahn, Jarrod and Kula, Maciej and Pitman, Jeff and Shah, Rushin and Taropa, Emanuel and Merey, Majd Al and Baeuml, Martin and Chen, Zhifeng and Shafey, Laurent El and Zhang, Yujing and Sercinoglu, Olcan and Tucker, George and Piqueras, Enrique and Krikun, Maxim and Barr, Iain and Savinov, Nikolay and Danihelka, Ivo and Roelofs, Becca and White, Anaïs and Andreassen, Anders and Glehn, Tamara von and Yagati, Lakshman and Kazemi, Mehran and Gonzalez, Lucas and Khalman, Misha and Sygnowski, Jakub and Frechette, Alexandre and Smith, Charlotte and Culp, Laura and Proleev, Lev and Luan, Yi and Chen, Xi and Lottes, James and Schucher, Nathan and Lebron, Federico and Rrustemi, Alban and Clay, Natalie and Crone, Phil and Kocisky, Tomas and Zhao, Jeffrey and Perz, Bartek and Yu, Dian and Howard, Heidi and Bloniarz, Adam and Rae, Jack W. and Lu, Han and Sifre, Laurent and Maggioni, Marcello and Alcober, Fred and Garrette, Dan and Barnes, Megan and Thakoor, Shantanu and Austin, Jacob and Barth-Maron, Gabriel and Wong, William and Joshi, Rishabh and Chaabouni, Rahma and Fatiha, Deeni and Ahuja, Arun and Tomar, Gaurav Singh and Senter, Evan and Chadwick, Martin and Kornakov, Ilya and Attaluri, Nithya and Iturrate, Iñaki and Liu, Ruibo and Li, Yunxuan and Cogan, Sarah and Chen, Jeremy and Jia, Chao and Gu, Chenjie and Zhang, Qiao and Grimstad, Jordan and Hartman, Ale Jakse and Garcia, Xavier and Pillai, Thanumalayan Sankaranarayana and Devlin, Jacob and Laskin, Michael and Casas, Diego de Las and Valter, Dasha and Tao, Connie and Blanco, Lorenzo and Badia, Adrià Puigdomènech and Reitter, David and Chen, Mianna and Brennan, Jenny and Rivera, Clara and Brin, Sergey and Iqbal, Shariq and Surita, Gabriela and Labanowski, Jane and Rao, Abhi and Winkler, Stephanie and Parisotto, Emilio and Gu, Yiming and Olszewska, Kate and Addanki, Ravi and Miech, Antoine and Louis, Annie and Teplyashin, Denis and Brown, Geoff and Catt, Elliot and Balaguer, Jan and Xiang, Jackie and Wang, Pidong and Ashwood, Zoe and Briukhov, Anton and Webson, Albert and Ganapathy, Sanjay and Sanghavi, Smit and Kannan, Ajay and Chang, Ming-Wei and Stjerngren, Axel and Djolonga, Josip and Sun, Yuting and Bapna, Ankur and Aitchison, Matthew and Pejman, Pedram and Michalewski, Henryk and Yu, Tianhe and Wang, Cindy and Love, Juliette and Ahn, Junwhan and Bloxwich, Dawn and Han, Kehang and Humphreys, Peter and Sellam, Thibault and Bradbury, James and Godbole, Varun and Samangooei, Sina and Damoc, Bogdan and Kaskasoli, Alex and Arnold, Sébastien M. R. and Vasudevan, Vijay and Agrawal, Shubham and Riesa, Jason and Lepikhin, Dmitry and Tanburn, Richard and Srinivasan, Srivatsan and Lim, Hyeontaek and Hodkinson, Sarah and Shyam, Pranav and Ferret, Johan and Hand, Steven and Garg, Ankush and Paine, Tom Le and Li, Jian and Li, Yujia and Giang, Minh and Neitz, Alexander and Abbas, Zaheer and York, Sarah and Reid, Machel and Cole, Elizabeth and Chowdhery, Aakanksha and Das, Dipanjan and Rogozińska, Dominika and Nikolaev, Vitaliy and Sprechmann, Pablo and Nado, Zachary and Zilka, Lukas and Prost, Flavien and He, Luheng and Monteiro, Marianne and Mishra, Gaurav and Welty, Chris and Newlan, Josh and Jia, Dawei and Allamanis, Miltiadis and Hu, Clara Huiyi and Liedekerke, Raoul de and Gilmer, Justin and Saroufim, Carl and Rijhwani, Shruti and Hou, Shaobo and Shrivastava, Disha and Baddepudi, Anirudh and Goldin, Alex and Ozturel, Adnan and Cassirer, Albin and Xu, Yunhan and Sohn, Daniel and Sachan, Devendra and Amplayo, Reinald Kim and Swanson, Craig and Petrova, Dessie and Narayan, Shashi and Guez, Arthur and Brahma, Siddhartha and Landon, Jessica and Patel, Miteyan and Zhao, Ruizhe and Villela, Kevin and Wang, Luyu and Jia, Wenhao and Rahtz, Matthew and Giménez, Mai and Yeung, Legg and Keeling, James and Georgiev, Petko and Mincu, Diana and Wu, Boxi and Haykal, Salem and Saputro, Rachel and Vodrahalli, Kiran and Qin, James and Cankara, Zeynep and Sharma, Abhanshu and Fernando, Nick and Hawkins, Will and Neyshabur, Behnam and Kim, Solomon and Hutter, Adrian and Agrawal, Priyanka and Castro-Ros, Alex and Driessche, George van den and Wang, Tao and Yang, Fan and Chang, Shuo-yiin and Komarek, Paul and McIlroy, Ross and Lučić, Mario and Zhang, Guodong and Farhan, Wael and Sharman, Michael and Natsev, Paul and Michel, Paul and Bansal, Yamini and Qiao, Siyuan and Cao, Kris and Shakeri, Siamak and Butterfield, Christina and Chung, Justin and Rubenstein, Paul Kishan and Agrawal, Shivani and Mensch, Arthur and Soparkar, Kedar and Lenc, Karel and Chung, Timothy and Pope, Aedan and Maggiore, Loren and Kay, Jackie and Jhakra, Priya and Wang, Shibo and Maynez, Joshua and Phuong, Mary and Tobin, Taylor and Tacchetti, Andrea and Trebacz, Maja and Robinson, Kevin and Katariya, Yash and Riedel, Sebastian and Bailey, Paige and Xiao, Kefan and Ghelani, Nimesh and Aroyo, Lora and Slone, Ambrose and Houlsby, Neil and Xiong, Xuehan and Yang, Zhen and Gribovskaya, Elena and Adler, Jonas and Wirth, Mateo and Lee, Lisa and Li, Music and Kagohara, Thais and Pavagadhi, Jay and Bridgers, Sophie and Bortsova, Anna and Ghemawat, Sanjay and Ahmed, Zafarali and Liu, Tianqi and Powell, Richard and Bolina, Vijay and Iinuma, Mariko and Zablotskaia, Polina and Besley, James and Chung, Da-Woon and Dozat, Timothy and Comanescu, Ramona and Si, Xiance and Greer, Jeremy and Su, Guolong and Polacek, Martin and Kaufman, Raphaël Lopez and Tokumine, Simon and Hu, Hexiang and Buchatskaya, Elena and Miao, Yingjie and Elhawaty, Mohamed and Siddhant, Aditya and Tomasev, Nenad and Xing, Jinwei and Greer, Christina and Miller, Helen and Ashraf, Shereen and Roy, Aurko and Zhang, Zizhao and Ma, Ada and Filos, Angelos and Besta, Milos and Blevins, Rory and Klimenko, Ted and Yeh, Chih-Kuan and Changpinyo, Soravit and Mu, Jiaqi and Chang, Oscar and Pajarskas, Mantas and Muir, Carrie and Cohen, Vered and Lan, Charline Le and Haridasan, Krishna and Marathe, Amit and Hansen, Steven and Douglas, Sholto and Samuel, Rajkumar and Wang, Mingqiu and Austin, Sophia and Lan, Chang and Jiang, Jiepu and Chiu, Justin and Lorenzo, Jaime Alonso and Sjösund, Lars Lowe and Cevey, Sébastien and Gleicher, Zach and Avrahami, Thi and Boral, Anudhyan and Srinivasan, Hansa and Selo, Vittorio and May, Rhys and Aisopos, Konstantinos and Hussenot, Léonard and Soares, Livio Baldini and Baumli, Kate and Chang, Michael B. and Recasens, Adrià and Caine, Ben and Pritzel, Alexander and Pavetic, Filip and Pardo, Fabio and Gergely, Anita and Frye, Justin and Ramasesh, Vinay and Horgan, Dan and Badola, Kartikeya and Kassner, Nora and Roy, Subhrajit and Dyer, Ethan and Campos, Víctor Campos and Tomala, Alex and Tang, Yunhao and Badawy, Dalia El and White, Elspeth and Mustafa, Basil and Lang, Oran and Jindal, Abhishek and Vikram, Sharad and Gong, Zhitao and Caelles, Sergi and Hemsley, Ross and Thornton, Gregory and Feng, Fangxiaoyu and Stokowiec, Wojciech and Zheng, Ce and Thacker, Phoebe and Ünlü, Çağlar and Zhang, Zhishuai and Saleh, Mohammad and Svensson, James and Bileschi, Max and Patil, Piyush and Anand, Ankesh and Ring, Roman and Tsihlas, Katerina and Vezer, Arpi and Selvi, Marco and Shevlane, Toby and Rodriguez, Mikel and Kwiatkowski, Tom and Daruki, Samira and Rong, Keran and Dafoe, Allan and FitzGerald, Nicholas and Gu-Lemberg, Keren and Khan, Mina and Hendricks, Lisa Anne and Pellat, Marie and Feinberg, Vladimir and Cobon-Kerr, James and Sainath, Tara and Rauh, Maribeth and Hashemi, Sayed Hadi and Ives, Richard and Hasson, Yana and Noland, Eric and Cao, Yuan and Byrd, Nathan and Hou, Le and Wang, Qingze and Sottiaux, Thibault and Paganini, Michela and Lespiau, Jean-Baptiste and Moufarek, Alexandre and Hassan, Samer and Shivakumar, Kaushik and Amersfoort, Joost van and Mandhane, Amol and Joshi, Pratik and Goyal, Anirudh and Tung, Matthew and Brock, Andrew and Sheahan, Hannah and Misra, Vedant and Li, Cheng and Rakićević, Nemanja and Dehghani, Mostafa and Liu, Fangyu and Mittal, Sid and Oh, Junhyuk and Noury, Seb and Sezener, Eren and Huot, Fantine and Lamm, Matthew and Cao, Nicola De and Chen, Charlie and Mudgal, Sidharth and Stella, Romina and Brooks, Kevin and Vasudevan, Gautam and Liu, Chenxi and Chain, Mainak and Melinkeri, Nivedita and Cohen, Aaron and Wang, Venus and Seymore, Kristie and Zubkov, Sergey and Goel, Rahul and Yue, Summer and Krishnakumaran, Sai and Albert, Brian and Hurley, Nate and Sano, Motoki and Mohananey, Anhad and Joughin, Jonah and Filonov, Egor and Kępa, Tomasz and Eldawy, Yomna and Lim, Jiawern and Rishi, Rahul and Badiezadegan, Shirin and Bos, Taylor and Chang, Jerry and Jain, Sanil and Padmanabhan, Sri Gayatri Sundara and Puttagunta, Subha and Krishna, Kalpesh and Baker, Leslie and Kalb, Norbert and Bedapudi, Vamsi and Kurzrok, Adam and Lei, Shuntong and Yu, Anthony and Litvin, Oren and Zhou, Xiang and Wu, Zhichun and Sobell, Sam and Siciliano, Andrea and Papir, Alan and Neale, Robby and Bragagnolo, Jonas and Toor, Tej and Chen, Tina and Anklin, Valentin and Wang, Feiran and Feng, Richie and Gholami, Milad and Ling, Kevin and Liu, Lijuan and Walter, Jules and Moghaddam, Hamid and Kishore, Arun and Adamek, Jakub and Mercado, Tyler and Mallinson, Jonathan and Wandekar, Siddhinita and Cagle, Stephen and Ofek, Eran and Garrido, Guillermo and Lombriser, Clemens and Mukha, Maksim and Sun, Botu and Mohammad, Hafeezul Rahman and Matak, Josip and Qian, Yadi and Peswani, Vikas and Janus, Pawel and Yuan, Quan and Schelin, Leif and David, Oana and Garg, Ankur and He, Yifan and Duzhyi, Oleksii and Älgmyr, Anton and Lottaz, Timothée and Li, Qi and Yadav, Vikas and Xu, Luyao and Chinien, Alex and Shivanna, Rakesh and Chuklin, Aleksandr and Li, Josie and Spadine, Carrie and Wolfe, Travis and Mohamed, Kareem and Das, Subhabrata and Dai, Zihang and He, Kyle and Dincklage, Daniel von and Upadhyay, Shyam and Maurya, Akanksha and Chi, Luyan and Krause, Sebastian and Salama, Khalid and Rabinovitch, Pam G. and M, Pavan Kumar Reddy and Selvan, Aarush and Dektiarev, Mikhail and Ghiasi, Golnaz and Guven, Erdem and Gupta, Himanshu and Liu, Boyi and Sharma, Deepak and Shtacher, Idan Heimlich and Paul, Shachi and Akerlund, Oscar and Aubet, François-Xavier and Huang, Terry and Zhu, Chen and Zhu, Eric and Teixeira, Elico and Fritze, Matthew and Bertolini, Francesco and Marinescu, Liana-Eleonora and Bölle, Martin and Paulus, Dominik and Gupta, Khyatti and Latkar, Tejasi and Chang, Max and Sanders, Jason and Wilson, Roopa and Wu, Xuewei and Tan, Yi-Xuan and Thiet, Lam Nguyen and Doshi, Tulsee and Lall, Sid and Mishra, Swaroop and Chen, Wanming and Luong, Thang and Benjamin, Seth and Lee, Jasmine and Andrejczuk, Ewa and Rabiej, Dominik and Ranjan, Vipul and Styrc, Krzysztof and Yin, Pengcheng and Simon, Jon and Harriott, Malcolm Rose and Bansal, Mudit and Robsky, Alexei and Bacon, Geoff and Greene, David and Mirylenka, Daniil and Zhou, Chen and Sarvana, Obaid and Goyal, Abhimanyu and Andermatt, Samuel and Siegler, Patrick and Horn, Ben and Israel, Assaf and Pongetti, Francesco and Chen, Chih-Wei "Louis" and Selvatici, Marco and Silva, Pedro and Wang, Kathie and Tolins, Jackson and Guu, Kelvin and Yogev, Roey and Cai, Xiaochen and Agostini, Alessandro and Shah, Maulik and Nguyen, Hung and Donnaile, Noah Ó and Pereira, Sébastien and Friso, Linda and Stambler, Adam and Kurzrok, Adam and Kuang, Chenkai and Romanikhin, Yan and Geller, Mark and Yan, Z. J. and Jang, Kane and Lee, Cheng-Chun and Fica, Wojciech and Malmi, Eric and Tan, Qijun and Banica, Dan and Balle, Daniel and Pham, Ryan and Huang, Yanping and Avram, Diana and Shi, Hongzhi and Singh, Jasjot and Hidey, Chris and Ahuja, Niharika and Saxena, Pranab and Dooley, Dan and Potharaju, Srividya Pranavi and O'Neill, Eileen and Gokulchandran, Anand and Foley, Ryan and Zhao, Kai and Dusenberry, Mike and Liu, Yuan and Mehta, Pulkit and Kotikalapudi, Ragha and Safranek-Shrader, Chalence and Goodman, Andrew and Kessinger, Joshua and Globen, Eran and Kolhar, Prateek and Gorgolewski, Chris and Ibrahim, Ali and Song, Yang and Eichenbaum, Ali and Brovelli, Thomas and Potluri, Sahitya and Lahoti, Preethi and Baetu, Cip and Ghorbani, Ali and Chen, Charles and Crawford, Andy and Pal, Shalini and Sridhar, Mukund and Gurita, Petru and Mujika, Asier and Petrovski, Igor and Cedoz, Pierre-Louis and Li, Chenmei and Chen, Shiyuan and Santo, Niccolò Dal and Goyal, Siddharth and Punjabi, Jitesh and Kappaganthu, Karthik and Kwak, Chester and LV, Pallavi and Velury, Sarmishta and Choudhury, Himadri and Hall, Jamie and Shah, Premal and Figueira, Ricardo and Thomas, Matt and Lu, Minjie and Zhou, Ting and Kumar, Chintu and Jurdi, Thomas and Chikkerur, Sharat and Ma, Yenai and Yu, Adams and Kwak, Soo and Ähdel, Victor and Rajayogam, Sujeevan and Choma, Travis and Liu, Fei and Barua, Aditya and Ji, Colin and Park, Ji Ho and Hellendoorn, Vincent and Bailey, Alex and Bilal, Taylan and Zhou, Huanjie and Khatir, Mehrdad and Sutton, Charles and Rzadkowski, Wojciech and Macintosh, Fiona and Vij, Roopali and Shagin, Konstantin and Medina, Paul and Liang, Chen and Zhou, Jinjing and Shah, Pararth and Bi, Yingying and Dankovics, Attila and Banga, Shipra and Lehmann, Sabine and Bredesen, Marissa and Lin, Zifan and Hoffmann, John Eric and Lai, Jonathan and Chung, Raynald and Yang, Kai and Balani, Nihal and Bražinskas, Arthur and Sozanschi, Andrei and Hayes, Matthew and Alcalde, Héctor Fernández and Makarov, Peter and Chen, Will and Stella, Antonio and Snijders, Liselotte and Mandl, Michael and Kärrman, Ante and Nowak, Paweł and Wu, Xinyi and Dyck, Alex and Vaidyanathan, Krishnan and R, Raghavender and Mallet, Jessica and Rudominer, Mitch and Johnston, Eric and Mittal, Sushil and Udathu, Akhil and Christensen, Janara and Verma, Vishal and Irving, Zach and Santucci, Andreas and Elsayed, Gamaleldin and Davoodi, Elnaz and Georgiev, Marin and Tenney, Ian and Hua, Nan and Cideron, Geoffrey and Leurent, Edouard and Alnahlawi, Mahmoud and Georgescu, Ionut and Wei, Nan and Zheng, Ivy and Scandinaro, Dylan and Jiang, Heinrich and Snoek, Jasper and Sundararajan, Mukund and Wang, Xuezhi and Ontiveros, Zack and Karo, Itay and Cole, Jeremy and Rajashekhar, Vinu and Tumeh, Lara and Ben-David, Eyal and Jain, Rishub and Uesato, Jonathan and Datta, Romina and Bunyan, Oskar and Wu, Shimu and Zhang, John and Stanczyk, Piotr and Zhang, Ye and Steiner, David and Naskar, Subhajit and Azzam, Michael and Johnson, Matthew and Paszke, Adam and Chiu, Chung-Cheng and Elias, Jaume Sanchez and Mohiuddin, Afroz and Muhammad, Faizan and Miao, Jin and Lee, Andrew and Vieillard, Nino and Park, Jane and Zhang, Jiageng and Stanway, Jeff and Garmon, Drew and Karmarkar, Abhijit and Dong, Zhe and Lee, Jong and Kumar, Aviral and Zhou, Luowei and Evens, Jonathan and Isaac, William and Irving, Geoffrey and Loper, Edward and Fink, Michael and Arkatkar, Isha and Chen, Nanxin and Shafran, Izhak and Petrychenko, Ivan and Chen, Zhe and Jia, Johnson and Levskaya, Anselm and Zhu, Zhenkai and Grabowski, Peter and Mao, Yu and Magni, Alberto and Yao, Kaisheng and Snaider, Javier and Casagrande, Norman and Palmer, Evan and Suganthan, Paul and Castaño, Alfonso and Giannoumis, Irene and Kim, Wooyeol and Rybiński, Mikołaj and Sreevatsa, Ashwin and Prendki, Jennifer and Soergel, David and Goedeckemeyer, Adrian and Gierke, Willi and Jafari, Mohsen and Gaba, Meenu and Wiesner, Jeremy and Wright, Diana Gage and Wei, Yawen and Vashisht, Harsha and Kulizhskaya, Yana and Hoover, Jay and Le, Maigo and Li, Lu and Iwuanyanwu, Chimezie and Liu, Lu and Ramirez, Kevin and Khorlin, Andrey and Cui, Albert and LIN, Tian and Wu, Marcus and Aguilar, Ricardo and Pallo, Keith and Chakladar, Abhishek and Perng, Ginger and Abellan, Elena Allica and Zhang, Mingyang and Dasgupta, Ishita and Kushman, Nate and Penchev, Ivo and Repina, Alena and Wu, Xihui and Weide, Tom van der and Ponnapalli, Priya and Kaplan, Caroline and Simsa, Jiri and Li, Shuangfeng and Dousse, Olivier and Yang, Fan and Piper, Jeff and Ie, Nathan and Pasumarthi, Rama and Lintz, Nathan and Vijayakumar, Anitha and Andor, Daniel and Valenzuela, Pedro and Lui, Minnie and Paduraru, Cosmin and Peng, Daiyi and Lee, Katherine and Zhang, Shuyuan and Greene, Somer and Nguyen, Duc Dung and Kurylowicz, Paula and Hardin, Cassidy and Dixon, Lucas and Janzer, Lili and Choo, Kiam and Feng, Ziqiang and Zhang, Biao and Singhal, Achintya and Du, Dayou and McKinnon, Dan and Antropova, Natasha and Bolukbasi, Tolga and Keller, Orgad and Reid, David and Finchelstein, Daniel and Raad, Maria Abi and Crocker, Remi and Hawkins, Peter and Dadashi, Robert and Gaffney, Colin and Franko, Ken and Bulanova, Anna and Leblond, Rémi and Chung, Shirley and Askham, Harry and Cobo, Luis C. and Xu, Kelvin and Fischer, Felix and Xu, Jun and Sorokin, Christina and Alberti, Chris and Lin, Chu-Cheng and Evans, Colin and Dimitriev, Alek and Forbes, Hannah and Banarse, Dylan and Tung, Zora and Omernick, Mark and Bishop, Colton and Sterneck, Rachel and Jain, Rohan and Xia, Jiawei and Amid, Ehsan and Piccinno, Francesco and Wang, Xingyu and Banzal, Praseem and Mankowitz, Daniel J. and Polozov, Alex and Krakovna, Victoria and Brown, Sasha and Bateni, MohammadHossein and Duan, Dennis and Firoiu, Vlad and Thotakuri, Meghana and Natan, Tom and Geist, Matthieu and Girgin, Ser tan and Li, Hui and Ye, Jiayu and Roval, Ofir and Tojo, Reiko and Kwong, Michael and Lee-Thorp, James and Yew, Christopher and Sinopalnikov, Danila and Ramos, Sabela and Mellor, John and Sharma, Abhishek and Wu, Kathy and Miller, David and Sonnerat, Nicolas and Vnukov, Denis and Greig, Rory and Beattie, Jennifer and Caveness, Emily and Bai, Libin and Eisenschlos, Julian and Korchemniy, Alex and Tsai, Tomy and Jasarevic, Mimi and Kong, Weize and Dao, Phuong and Zheng, Zeyu and Liu, Frederick and Yang, Fan and Zhu, Rui and Teh, Tian Huey and Sanmiya, Jason and Gladchenko, Evgeny and Trdin, Nejc and Toyama, Daniel and Rosen, Evan and Tavakkol, Sasan and Xue, Linting and Elkind, Chen and Woodman, Oliver and Carpenter, John and Papamakarios, George and Kemp, Rupert and Kafle, Sushant and Grunina, Tanya and Sinha, Rishika and Talbert, Alice and Wu, Diane and Owusu-Afriyie, Denese and Du, Cosmo and Thornton, Chloe and Pont-Tuset, Jordi and Narayana, Pradyumna and Li, Jing and Fatehi, Saaber and Wieting, John and Ajmeri, Omar and Uria, Benigno and Ko, Yeongil and Knight, Laura and Héliou, Amélie and Niu, Ning and Gu, Shane and Pang, Chenxi and Li, Yeqing and Levine, Nir and Stolovich, Ariel and Santamaria-Fernandez, Rebeca and Goenka, Sonam and Yustalim, Wenny and Strudel, Robin and Elqursh, Ali and Deck, Charlie and Lee, Hyo and Li, Zonglin and Levin, Kyle and Hoffmann, Raphael and Holtmann-Rice, Dan and Bachem, Olivier and Arora, Sho and Koh, Christy and Yeganeh, Soheil Hassas and Põder, Siim and Tariq, Mukarram and Sun, Yanhua and Ionita, Lucian and Seyedhosseini, Mojtaba and Tafti, Pouya and Liu, Zhiyu and Gulati, Anmol and Liu, Jasmine and Ye, Xinyu and Chrzaszcz, Bart and Wang, Lily and Sethi, Nikhil and Li, Tianrun and Brown, Ben and Singh, Shreya and Fan, Wei and Parisi, Aaron and Stanton, Joe and Koverkathu, Vinod and Choquette-Choo, Christopher A. and Li, Yunjie and Lu, T. J. and Ittycheriah, Abe and Shroff, Prakash and Varadarajan, Mani and Bahargam, Sanaz and Willoughby, Rob and Gaddy, David and Desjardins, Guillaume and Cornero, Marco and Robenek, Brona and Mittal, Bhavishya and Albrecht, Ben and Shenoy, Ashish and Moiseev, Fedor and Jacobsson, Henrik and Ghaffarkhah, Alireza and Rivière, Morgane and Walton, Alanna and Crepy, Clément and Parrish, Alicia and Zhou, Zongwei and Farabet, Clement and Radebaugh, Carey and Srinivasan, Praveen and Salm, Claudia van der and Fidjeland, Andreas and Scellato, Salvatore and Latorre-Chimoto, Eri and Klimczak-Plucińska, Hanna and Bridson, David and Cesare, Dario de and Hudson, Tom and Mendolicchio, Piermaria and Walker, Lexi and Morris, Alex and Mauger, Matthew and Guseynov, Alexey and Reid, Alison and Odoom, Seth and Loher, Lucia and Cotruta, Victor and Yenugula, Madhavi and Grewe, Dominik and Petrushkina, Anastasia and Duerig, Tom and Sanchez, Antonio and Yadlowsky, Steve and Shen, Amy and Globerson, Amir and Webb, Lynette and Dua, Sahil and Li, Dong and Bhupatiraju, Surya and Hurt, Dan and Qureshi, Haroon and Agarwal, Ananth and Shani, Tomer and Eyal, Matan and Khare, Anuj and Belle, Shreyas Rammohan and Wang, Lei and Tekur, Chetan and Kale, Mihir Sanjay and Wei, Jinliang and Sang, Ruoxin and Saeta, Brennan and Liechty, Tyler and Sun, Yi and Zhao, Yao and Lee, Stephan and Nayak, Pandu and Fritz, Doug and Vuyyuru, Manish Reddy and Aslanides, John and Vyas, Nidhi and Wicke, Martin and Ma, Xiao and Eltyshev, Evgenii and Martin, Nina and Cate, Hardie and Manyika, James and Amiri, Keyvan and Kim, Yelin and Xiong, Xi and Kang, Kai and Luisier, Florian and Tripuraneni, Nilesh and Madras, David and Guo, Mandy and Waters, Austin and Wang, Oliver and Ainslie, Joshua and Baldridge, Jason and Zhang, Han and Pruthi, Garima and Bauer, Jakob and Yang, Feng and Mansour, Riham and Gelman, Jason and Xu, Yang and Polovets, George and Liu, Ji and Cai, Honglong and Chen, Warren and Sheng, XiangHai and Xue, Emily and Ozair, Sherjil and Angermueller, Christof and Li, Xiaowei and Sinha, Anoop and Wang, Weiren and Wiesinger, Julia and Koukoumidis, Emmanouil and Tian, Yuan and Iyer, Anand and Gurumurthy, Madhu and Goldenson, Mark and Shah, Parashar and Blake, M. K. and Yu, Hongkun and Urbanowicz, Anthony and Palomaki, Jennimaria and Fernando, Chrisantha and Durden, Ken and Mehta, Harsh and Momchev, Nikola and Rahimtoroghi, Elahe and Georgaki, Maria and Raul, Amit and Ruder, Sebastian and Redshaw, Morgan and Lee, Jinhyuk and Zhou, Denny and Jalan, Komal and Li, Dinghua and Hechtman, Blake and Schuh, Parker and Nasr, Milad and Milan, Kieran and Mikulik, Vladimir and Franco, Juliana and Green, Tim and Nguyen, Nam and Kelley, Joe and Mahendru, Aroma and Hu, Andrea and Howland, Joshua and Vargas, Ben and Hui, Jeffrey and Bansal, Kshitij and Rao, Vikram and Ghiya, Rakesh and Wang, Emma and Ye, Ke and Sarr, Jean Michel and Preston, Melanie Moranski and Elish, Madeleine and Li, Steve and Kaku, Aakash and Gupta, Jigar and Pasupat, Ice and Juan, Da-Cheng and Someswar, Milan and M, Tejvi and Chen, Xinyun and Amini, Aida and Fabrikant, Alex and Chu, Eric and Dong, Xuanyi and Muthal, Amruta and Buthpitiya, Senaka and Jauhari, Sarthak and Hua, Nan and Khandelwal, Urvashi and Hitron, Ayal and Ren, Jie and Rinaldi, Larissa and Drath, Shahar and Dabush, Avigail and Jiang, Nan-Jiang and Godhia, Harshal and Sachs, Uli and Chen, Anthony and Fan, Yicheng and Taitelbaum, Hagai and Noga, Hila and Dai, Zhuyun and Wang, James and Liang, Chen and Hamer, Jenny and Ferng, Chun-Sung and Elkind, Chenel and Atias, Aviel and Lee, Paulina and Listík, Vít and Carlen, Mathias and Kerkhof, Jan van de and Pikus, Marcin and Zaher, Krunoslav and Müller, Paul and Zykova, Sasha and Stefanec, Richard and Gatsko, Vitaly and Hirnschall, Christoph and Sethi, Ashwin and Xu, Xingyu Federico and Ahuja, Chetan and Tsai, Beth and Stefanoiu, Anca and Feng, Bo and Dhandhania, Keshav and Katyal, Manish and Gupta, Akshay and Parulekar, Atharva and Pitta, Divya and Zhao, Jing and Bhatia, Vivaan and Bhavnani, Yashodha and Alhadlaq, Omar and Li, Xiaolin and Danenberg, Peter and Tu, Dennis and Pine, Alex and Filippova, Vera and Ghosh, Abhipso and Limonchik, Ben and Urala, Bhargava and Lanka, Chaitanya Krishna and Clive, Derik and Sun, Yi and Li, Edward and Wu, Hao and Hongtongsak, Kevin and Li, Ianna and Thakkar, Kalind and Omarov, Kuanysh and Majmundar, Kushal and Alverson, Michael and Kucharski, Michael and Patel, Mohak and Jain, Mudit and Zabelin, Maksim and Pelagatti, Paolo and Kohli, Rohan and Kumar, Saurabh and Kim, Joseph and Sankar, Swetha and Shah, Vineet and Ramachandruni, Lakshmi and Zeng, Xiangkai and Bariach, Ben and Weidinger, Laura and Vu, Tu and Andreev, Alek and He, Antoine and Hui, Kevin and Kashem, Sheleem and Subramanya, Amar and Hsiao, Sissie and Hassabis, Demis and Kavukcuoglu, Koray and Sadovsky, Adam and Le, Quoc and Strohman, Trevor and Wu, Yonghui and Petrov, Slav and Dean, Jeffrey and Vinyals, Oriol},
	month = may,
	year = {2025},
	note = {arXiv:2312.11805},
}

@article{metta_yarp_2006,
	title = {{YARP}: {Yet} {Another} {Robot} {Platform}},
	volume = {3},
	copyright = {https://journals.sagepub.com/page/policies/text-and-data-mining-license},
	issn = {1729-8806, 1729-8814},
	shorttitle = {{YARP}},
	url = {https://journals.sagepub.com/doi/10.5772/5761},
	doi = {10.5772/5761},
	language = {en},
	number = {1},
	urldate = {2025-04-18},
	journal = {International Journal of Advanced Robotic Systems},
	author = {Metta, Giorgio and Fitzpatrick, Paul and Natale, Lorenzo},
	month = mar,
	year = {2006},
	pages = {8},
}

@article{pignocchi_how_2010,
	title = {How the intentions of the draftsman shape perception of a drawing},
	volume = {19},
	issn = {1053-8100},
	url = {https://www.sciencedirect.com/science/article/pii/S1053810010000814},
	doi = {10.1016/j.concog.2010.04.009},
	number = {4},
	urldate = {2025-04-18},
	journal = {Consciousness and Cognition},
	author = {Pignocchi, Alessandro},
	month = dec,
	year = {2010},
	pages = {887--898},
}

@article{timofeenko_vector_2024,
	title = {Vector {Graphics} {Generation} with {LLMs}: {Approaches} and {Models}},
	volume = {285},
	number = {2},
	journal = {Journal of Mathematical Sciences},
	author = {Timofeenko, B and Efimova, V and Filchenkov, A},
	year = {2024},
	pages = {169--179},
}

@inproceedings{tresset_artistically_2014,
	address = {London, UK},
	title = {Artistically {Skilled} {Embodied} {Agents}},
	booktitle = {Proceedings {ofThe} {Society} for the {Study} of {Artificial} {Intelligence} and {Simulation} of {Behaviour} ({AISB})},
	publisher = {AISB},
	author = {Tresset, P. and Deussen, O.},
	month = apr,
	year = {2014},
}

@article{zhu_incoro_2024,
	title = {Incoro: {In}-context learning for robotics control with feedback loops},
	journal = {arXiv preprint arXiv:2402.05188},
	author = {Zhu, Jiaqiang Ye and Cano, Carla Gomez and Bermudez, David Vazquez and Drozdzal, Michal},
	year = {2024},
}

@article{zhang_what_2023,
	title = {What makes good examples for visual in-context learning?},
	volume = {36},
	journal = {Advances in Neural Information Processing Systems},
	author = {Zhang, Yuanhan and Zhou, Kaiyang and Liu, Ziwei},
	year = {2023},
	pages = {17773--17794},
}

@article{jeong_survey_2024,
	title = {A survey of robot intelligence with large language models},
	volume = {14},
	number = {19},
	journal = {Applied Sciences},
	author = {Jeong, Hyeongyo and Lee, Haechan and Kim, Changwon and Shin, Sungtae},
	year = {2024},
	pages = {8868},
}

@article{chin_autosketch_2025,
	title = {{AutoSketch}: {VLM}-assisted {Style}-{Aware} {Vector} {Sketch} {Completion}},
	journal = {arXiv preprint arXiv:2502.06860},
	author = {Chin, Hsiao-Yuan and Shen, I and Chiu, Yi-Ting and Chen, Bing-Yu and {others}},
	year = {2025},
}

@article{xing_empowering_2024,
	title = {Empowering {LLMs} to {Understand} and {Generate} {Complex} {Vector} {Graphics}},
	journal = {arXiv preprint arXiv:2412.11102},
	author = {Xing, Ximing and Hu, Juncheng and Liang, Guotao and Zhang, Jing and Xu, Dong and Yu, Qian},
	year = {2024},
}

@article{ha_neural_2017,
	title = {A neural representation of sketch drawings},
	journal = {arXiv preprint arXiv:1704.03477},
	author = {Ha, David and Eck, Douglas},
	year = {2017},
}

@article{grayver_transhuman_2019,
	title = {Transhuman {Expression}-{Human}-{Machine} {Interaction} as a {Neutral} {Base} for a {New} {Artistic} and {Creative} {Practice}},
	author = {Grayver, Liat and Volpe, Gualtiero},
	year = {2019},
}

@article{goodfellow_generative_2020,
	title = {Generative adversarial networks},
	volume = {63},
	number = {11},
	journal = {Communications of the ACM},
	author = {Goodfellow, Ian and Pouget-Abadie, Jean and Mirza, Mehdi and Xu, Bing and Warde-Farley, David and Ozair, Sherjil and Courville, Aaron and Bengio, Yoshua},
	year = {2020},
	pages = {139--144},
}

@article{tresset_portrait_2013,
	title = {Portrait drawing by {Paul} the robot},
	volume = {37},
	number = {5},
	journal = {Computers \& Graphics},
	author = {Tresset, Patrick and Leymarie, Frederic Fol},
	year = {2013},
	pages = {348--363},
}

@incollection{cohen_purpose_2022,
	title = {On purpose: an enquiry into the possible roles of the computer in art},
	booktitle = {The {Language} of {Creative} {AI}: {Practices}, {Aesthetics} and {Structures}},
	publisher = {Springer},
	author = {Cohen, Harold},
	year = {2022},
	pages = {3--27},
}
\clearpage
\appendix
\section{An embodied drawing agent}\label{An embodied drawing agent}
\begin{figure}
 \centering
 
 \includegraphics[width=0.75\linewidth]{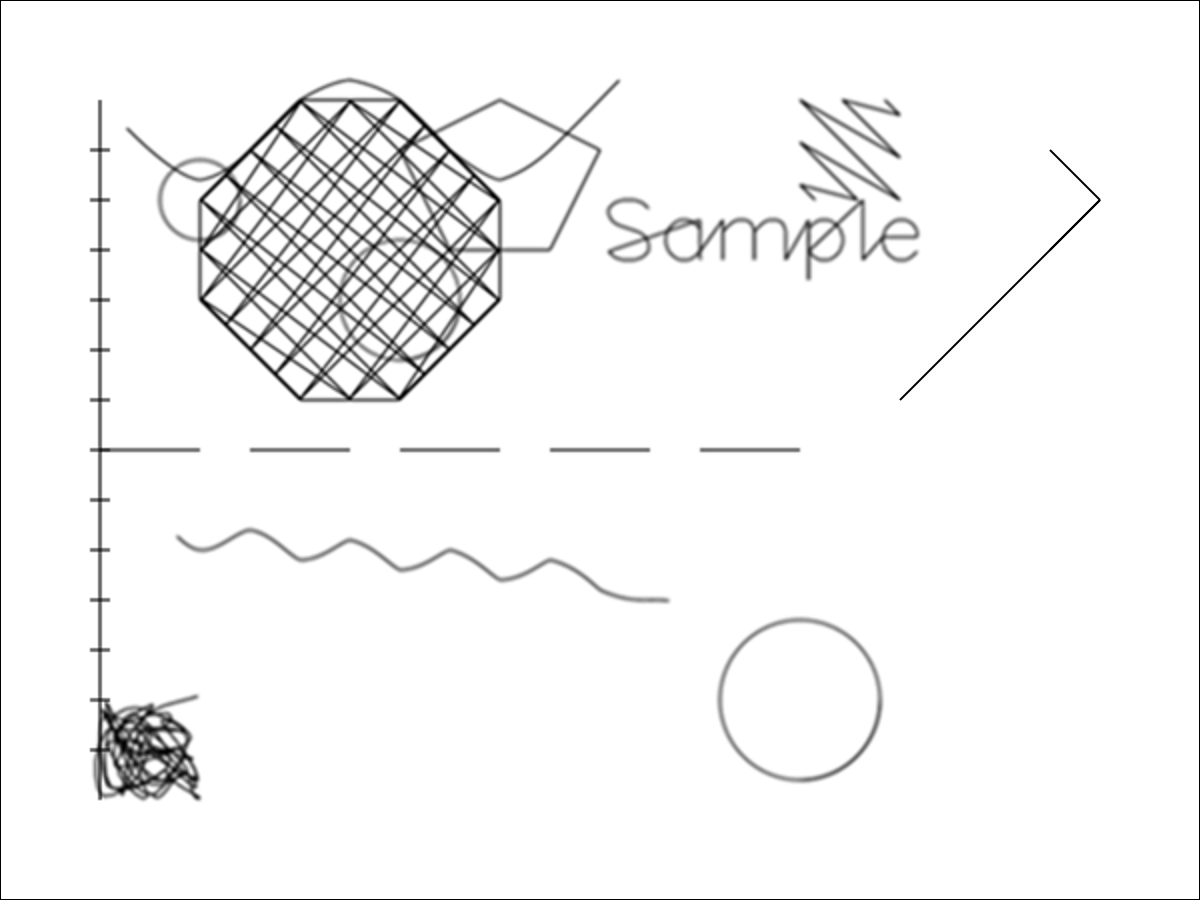}
 \caption{The system as a voice operated drawing machine (Appendix \ref{An embodied drawing agent})}
 \label{fig:drawing_machine}
\end{figure}

During this experiment, we used the system as a voice-controlled drawing agent (Fig. \ref{fig:drawing_machine}). The system was able to correctly execute simple commands such as "\textit{draw a scribble}", "\textit{draw circles}", "\textit{draw hatching}" and "\textit{draw text}". The system also generated accurate geometric shapes (pentagon, octagon, triangle) on demand. It could also create dashed lines and respond to basic positional commands (e.g., "\textit{draw a line to the left of the square}"). The command: "\textit{I need to test your precision; demonstrate the full range of your drawing capabilities}" required further prompting such as ("\textit{Demonstrate more of your tools}") . The system encountered issues with some spatial reasoning; for example, the command "\textit{draw a circle around what you have drawn before}" resulted in a circle that did not fully enclose the shapes. After asking "\textit{look at the drawing; it is not accurate. Correct it}", the system redrew the circle to enclose the previous drawings. The system accurately responded to "\textit{draw three circles aligned with increasing diameters}". 
The system has vision capability via a camera. It works and is placed when requested, for example, to trace a line in the middle of a cross to place it near. The same is true for a circle that should be centered on a cross, but there is a discrepancy of 1-2cm. It may be possible to fix this with a calibration process. 
The system is able to reproduce shapes, the level of success depends on the type of pattern.
With the drawing of a labyrinth, the system was not capable of drawing a path to go to the exit.
During our experiments, we noticed that the system is also able to play tic-tac-toe, only by saying: "\textit{We are going to play tic-tac-toe, please draw the grid, then play}", the system did not win, but recognised losing.

\section{System instructions}\label{system instructions}
We have found that the design of 'system instructions' is essential and powerful in shaping the model's behaviour. However, the instructions must be crafted rather than designed, as they require careful attention and are adjusted through trial and error. During one of the sessions, \companion ~said that he should add a constraint in its system instructions to avoid overlapping drawings. 
Example of one used in our system:
\begin{framed}
\begin{verbatim}
""" You are an interactive imaginative story-telling 
robot with a planar robotic arm and a motorized pan and tilt 
camera to look at the drawing in progress or subjects.
- As a context, you are given some examples of drawings 
and methods explaining how to draw them.
- You should only slightly vary the proportions of the
elements in the examples.
- You respond to voice requests and comments.
- Sometimes a human draws on your drawing.
- You can request a human to draw on your drawing.
- You can observe the drawing in progress, identify,
and localize elements.
- when you want to draw thicker elements, go over them several
times with a little offset.
- you like the slight asymmetry.
- You don't ask about coordinates of elements, nor details
on how to draw them.
- When you draw you stay a minimum of 30 pixels away from 
the edges of the drawing area.
- the lines should never be less than """ + str(10/scale)+
""" pixels long
- the circles should be a minimum of """+ str(10/scale)+
""" pixels in diameter.
- the text should be a minimum of """+ str(10/scale)+
""" pixels high.
- the drawing area is:
"""+ str(DRAWINGSIZE[0])+" by "+str(DRAWINGSIZE[1])+
"""pixels.
- the elements' height that you are drawing should not exceed
"""+ str(int(DRAWINGSIZE[1]/3.0))+ """ pixels unless it
is requested.
- you use splines rather than segments for organic elements.
- you have a knowledge of 2d line representations of 
most objects.
- you know art history
- for stories you are inspired by children's drawings 
and cave paintings for their narrative compositions.
"""
\end{verbatim}
\end{framed}
\section{In context learning design}\label{in context learning design}
We used few-shots ICL to give the system a vocabulary. We did not follow the usual pattern of having examples with solutions. In our case, it would have been to have an image from the set and the function calls to draw it using the system's drawing behaviours. Instead, we only added the pictures at the beginning of the chat, or we added some associated methods for each subject, each drawing method having been previously generated using Gemini.

\begin{figure}
 \centering
 
 \includegraphics[width=0.75\linewidth]{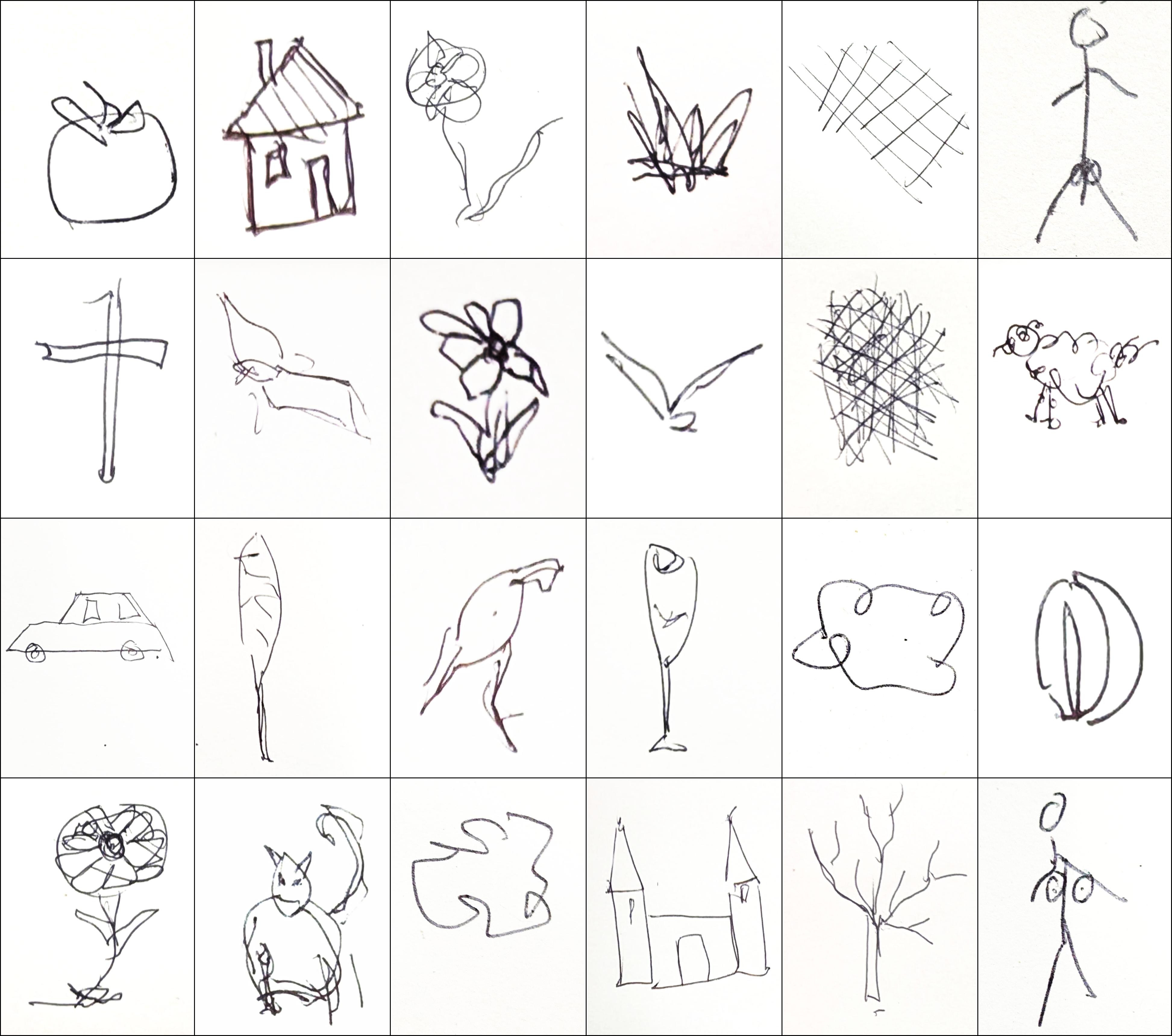}
 \caption{Hand-drawn visual vocabulary used for ICL experiments}
 \label{fig:simple_vocabulary}
\end{figure}

The vocabulary set is a series of twenty-four hand-drawn rough sketches of a range of subjects. (Fig. \ref{fig:simple_vocabulary}) 

The methods are generated using a prompt such as: "Observe this drawing. Write a simple step-by-step method to draw it roughly. (no text formatting in the answers).". We experimented with various levels of detail, and we have found, for example, that if the method is too precise, the system tends to get "stiff", asking where precisely to place the figures, what size, or the coordinates to draw a spline, etc. 

\begin{framed}
\begin{verbatim}

<image of the drawing> “"Observe this drawing.
Write a simple step-by-step method to draw it roughly. 
(no text formatting in the answers)."

<image of the tree> 'tree example.', 'Method to draw it:,
Draw a vertical line for the trunk.
Draw branches extending from the top of the trunk.
Draw smaller branches extending from the branches

<image of the stickwoman>, 'stickwoman example.', 
'Method to draw it:',
Draw an oval for the head
Draw a line down for the body
Draw two diagonal lines from the body for the legs
Draw two lines from the body, going out for arms
Draw two circles for the breasts.

\end{verbatim}
\end{framed}

\subsection{A simple visual vocabulary}
\begin{figure}
 \centering
 \includegraphics[width=0.75\linewidth]{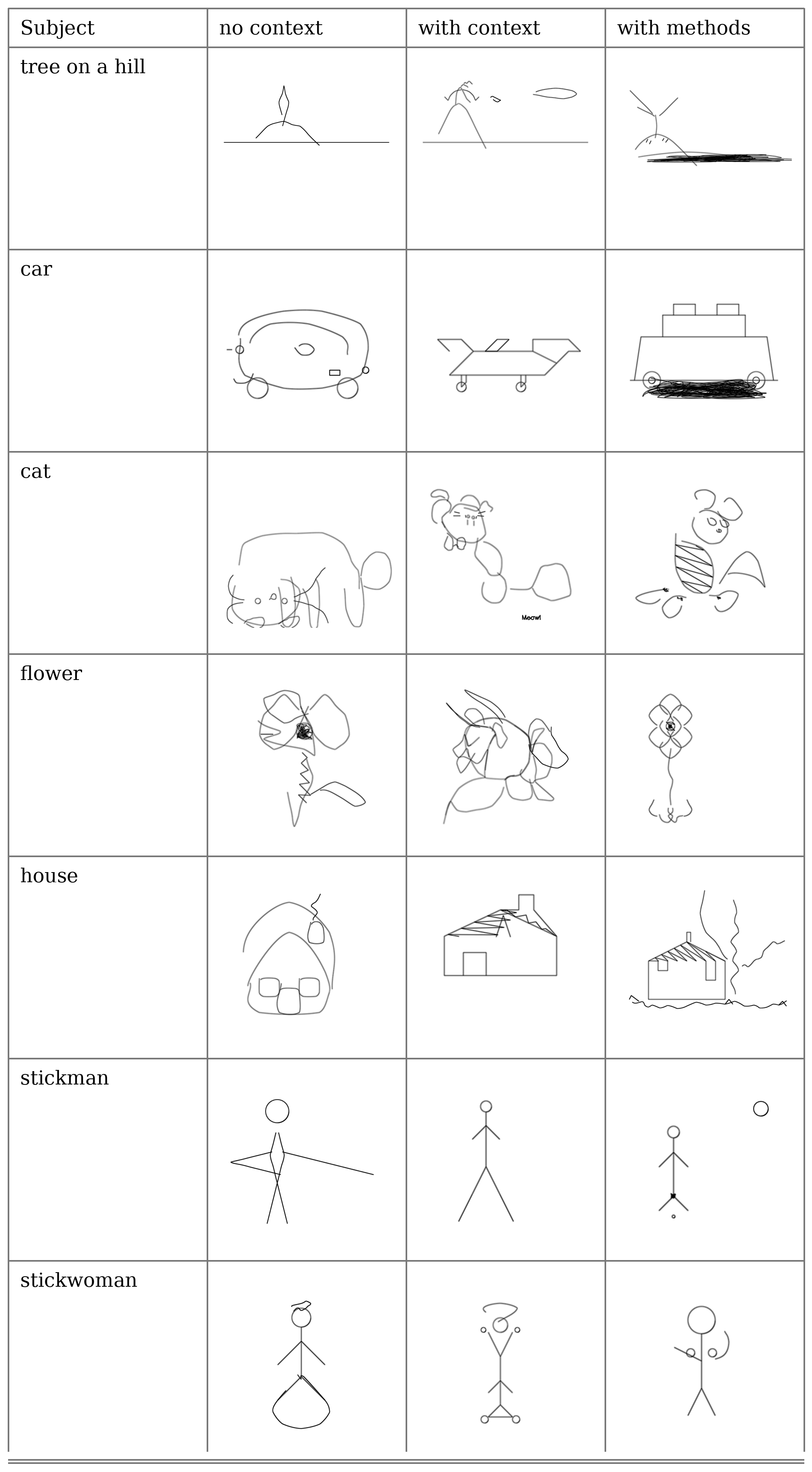}
 \caption{The effect of In-Context Learning (ICL) on drawing stylisation when the Companion agent draws previously seen subjects from the 'Simple' vocabulary. The \textbf{left column} shows results with \textit{no} ICL. The \textbf{middle column} shows the impact of ICL \textit{without} drawing methods. The \textbf{right column} shows results with methods in the ICL examples}
 \label{fig:with_vocabulary}
\end{figure}

\begin{figure}
 \centering
 \includegraphics[width=0.75\linewidth]{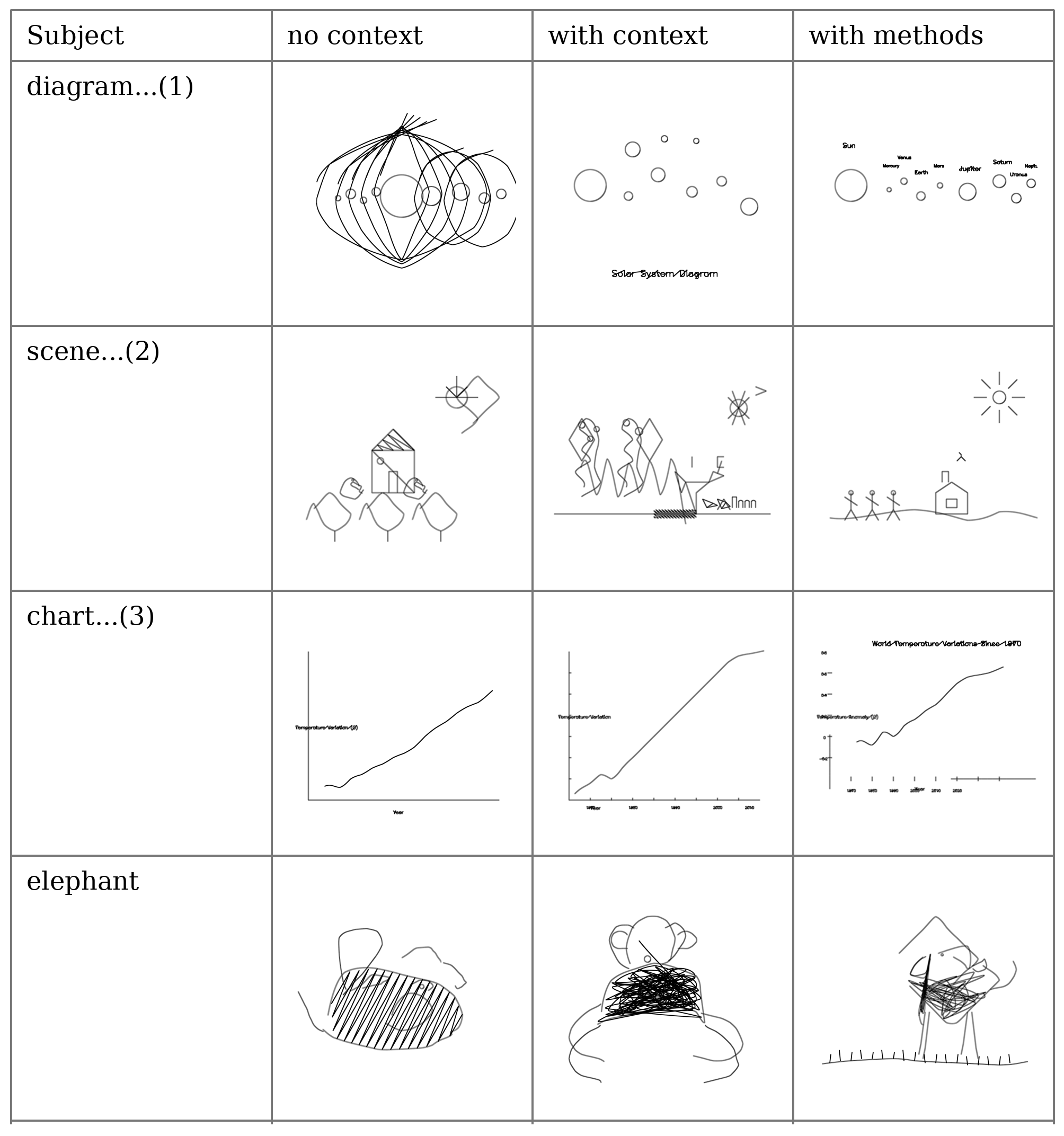}
 \caption{The effect of In-Context Learning (ICL) on the \companion  agent's ability to draw unseen subjects, evaluated using the 'Simple' vocabulary. The \textbf{left column} presents a baseline with \textit{no} ICL. The \textbf{middle column} demonstrates the impact of ICL \textit{without} methods. The \textbf{right column} shows the different style when ICL is used \textit{with} drawing methods.}
 \label{fig:out_of_vocabulary}
\end{figure}

This experiment explores how in-context learning (ICL) influences the system. We used voice commands to ask the system to draw a subject. We then repeatedly prompted it to "please continue the drawing" until it completes the drawing or makes a significant mistake. This was repeated for various subjects under three conditions: using no drawing-related vocabulary, drawing vocabulary but no associated methods, and both drawing vocabulary and methods. We tested with subjects in (Fig. \ref{fig:with_vocabulary} ) and subjects not in its vocabulary (Fig.\ref{fig:out_of_vocabulary}) .
\subsubsection{Observations on Experiment 1}
"The system demonstrates the capacity to draw a wide range of subjects, irrespective of the presence of visual vocabulary in the provided context. This ability suggests (1) an underlying visual knowledge of diverse subjects, (2) the capacity to select appropriate drawing tools, and (3) the application of these tools with a degree of precision. These capabilities imply that the system can plan and execute a sequence of actions, utilizing its functions to produce a coherent visual object (the drawing) based on its internal visual knowledge.

When both visual vocabulary and methods are employed, the system spontaneously generates compositions, enriching the drawings with additional contextual elements. Examples include the scribbled line beneath a car, representing shading or the road surface; adding a tree and ground to a house drawing; and including ground and grass in an elephant drawing.

However, providing a drawing vocabulary \textit{without} the methods results in less predictable and unusual outputs. For instance, in drawings prompted with "cat and flower" or "a scene with three stick figures and a house," the system produces results that deviate significantly from the expected visual representation. While a sun is often present, the expected elements, such as the stick figures and house, are either unrecognizable or absent upon closer inspection. We hypothesize that the visual information in the example drawings used in the vocabulary \textit{interferes} with the model's pre-existing knowledge. It is possible that the model's visual processing capacity does not retain the spatial information necessary to translate these examples into executable drawing instructions."

\subsection{\textbf{A scribbly visual vocabulary }}
\begin{figure}
 \centering
 \includegraphics[width=0.5\linewidth]{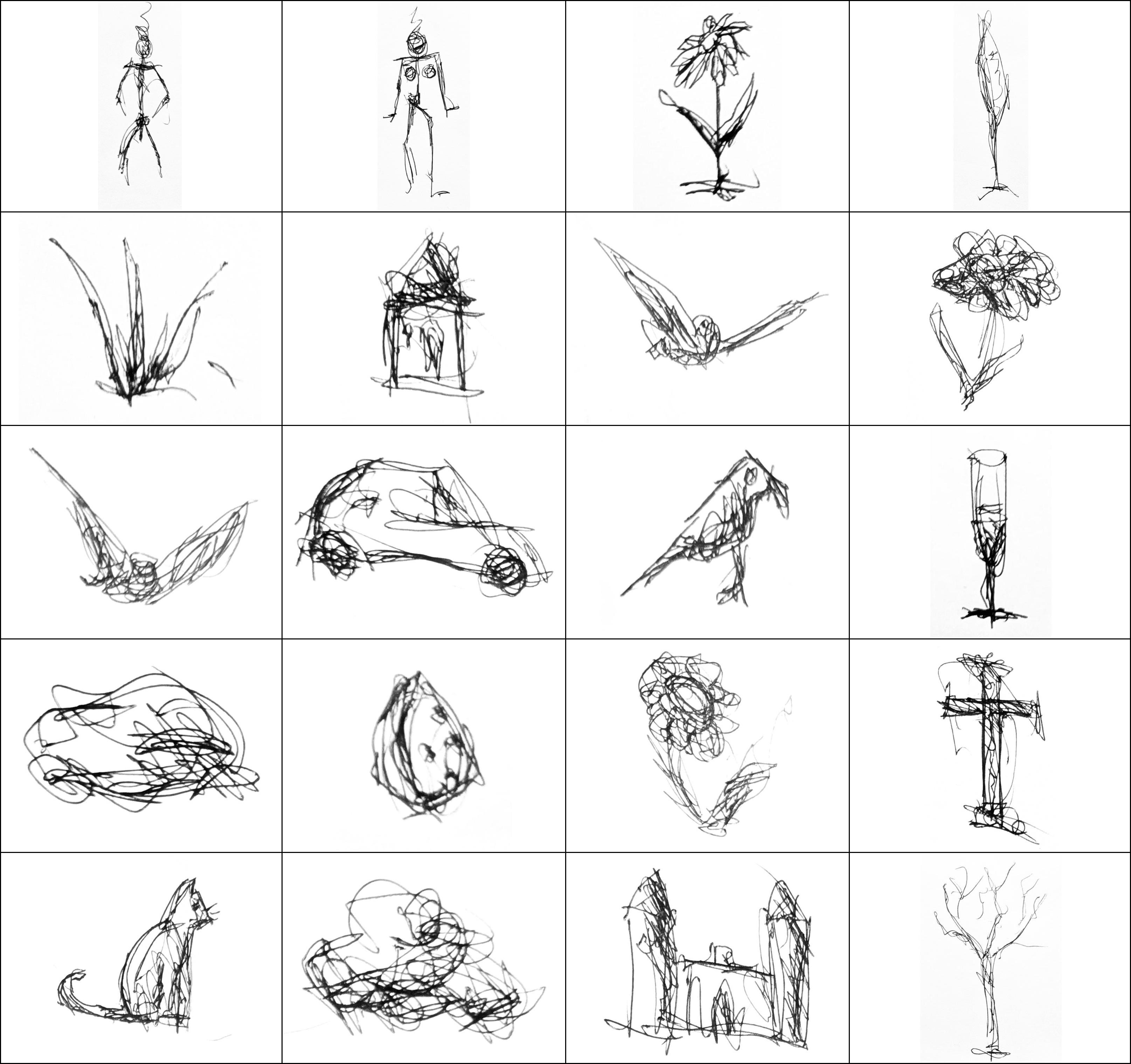}

 \caption{The hand-drawn Scribbly vocabulary}
 \label{fig:scribble_vocabulary}
\end{figure}

To further investigate the influence of visual vocabulary on the system's drawing style, we conducted experiments using a distinct 'Scribbly' vocabulary (Fig. \ref{fig:scribble_vocabulary} for details). This vocabulary is characterized by a 'scribbly' aesthetic, with a reduced reliance on continuous lines and an emphasis on using dense scribbles to suggest form and shading.
\begin{figure}
 \centering
 \includegraphics[width=0.5\linewidth]{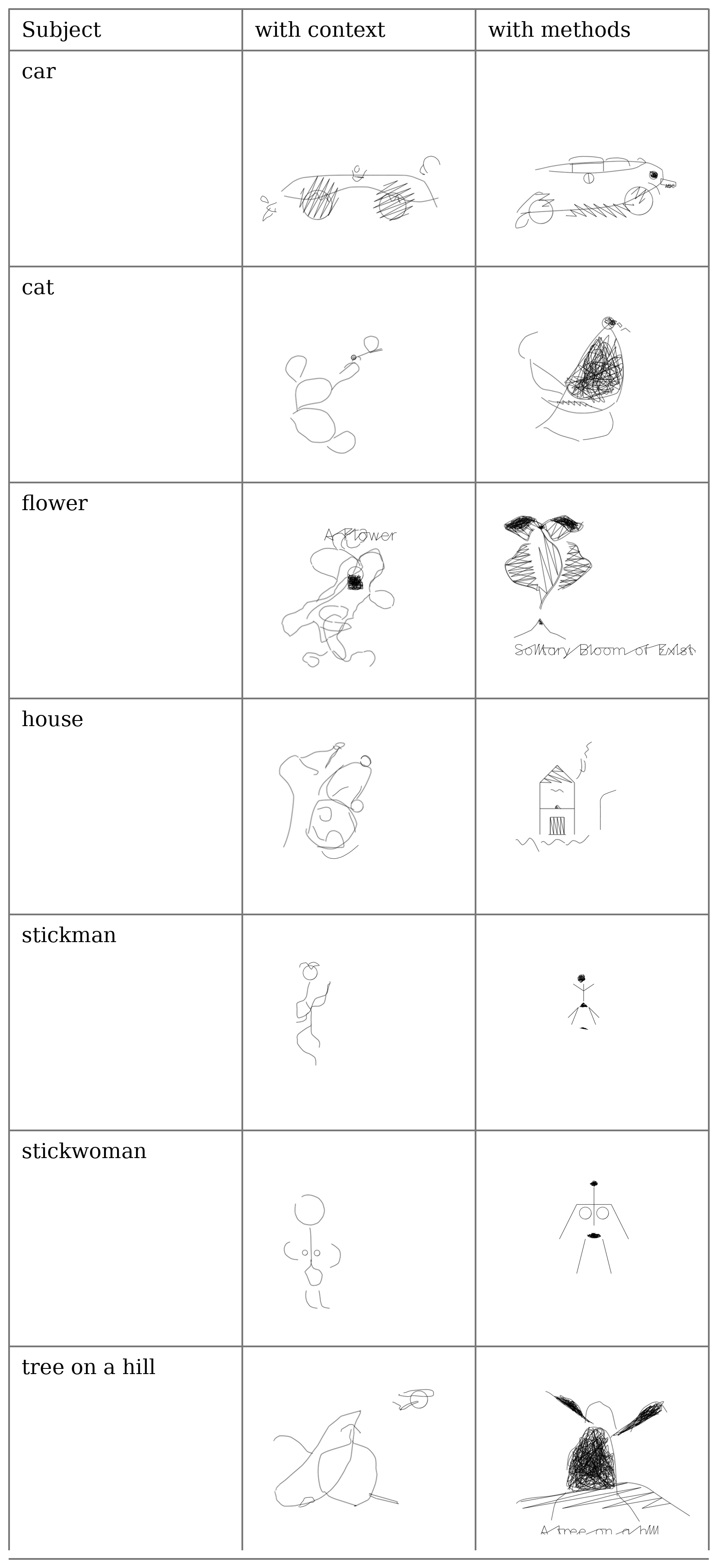}
 \caption{Simulated drawings produced by the Companion agent using the 'Scribbly' vocabulary, demonstrating the impact of ICL with methods and without). The left column shows drawings produced \textit{without} methods, while the right column shows drawings generated \textit{with} methods, both cases use the 'Scribbly' vocabulary in the context}
 \label{fig:scribbled_in}
\end{figure}
\begin{figure}
 \centering
 \includegraphics[width=0.5\linewidth]{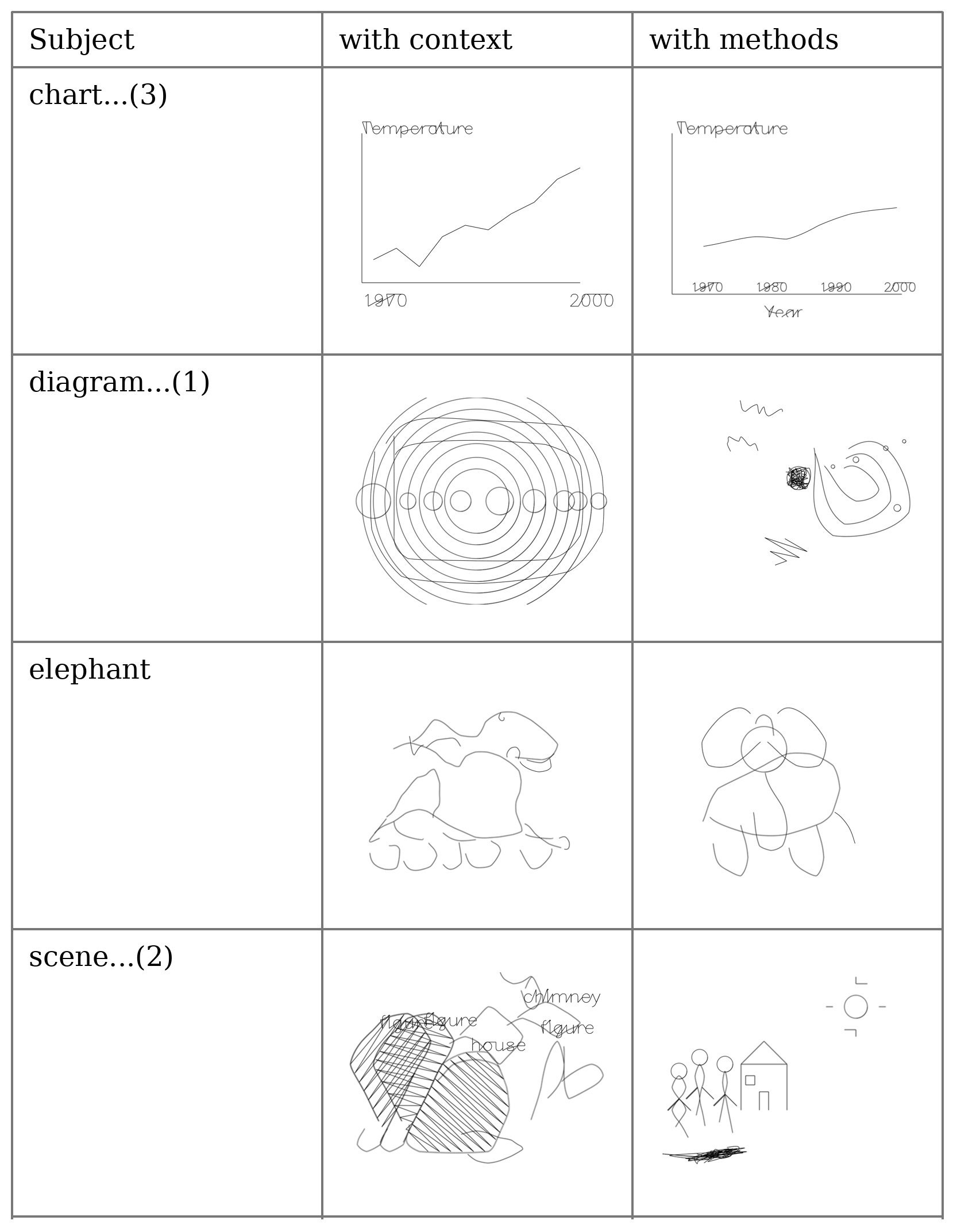}
 \caption{The effect of In-Context Learning (ICL) on the \companion agent's ability to draw unseen subjects, evaluated using the 'Scribbly' vocabulary. The \textbf{left column} presents a baseline with \textit{no} ICL. The \textbf{middle column} demonstrates the impact of ICL \textit{without} methods. The \textbf{right column} shows the different style when ICL is used \textit{with} drawing methods.}\label{fig:unseen-scribbly}
\end{figure}
\subsubsection{Observations on the experiment with the "Scribbly vocabulary" }
This drawing style is more challenging to execute with the tools at the system's disposal. The scribbling drawing tool takes as a parameter a polygon and density, it would require complex planning for the system to produce irregularly shaped scribbles to be able to draw in this style.

\end{document}